\documentstyle[11pt]{article}
\normalbaselines
\begin{document}
\bibliographystyle{unsrt}

\def\bea*{\begin{eqnarray*}}
\def\eea*{\end{eqnarray*}}
\def\ba{\begin{array}}
\def\ea{\end{array}}
% --------------------------------------------------------------
% If you want to see the names of the equations and references
% set below \count1=0 otherwise \count1=1
% --------------------------------------------------------------
\count1=1
% --------------------------------------------------------------
\def\be{\ifnum \count1=0 $$ \else \begin{equation}\fi}
\def\ee{\ifnum\count1=0 $$ \else \end{equation}\fi}
\def\ele(#1){\ifnum\count1=0 \eqno({\bf #1}) $$ \else
\label{#1}\end{equation}\fi}
\def\req(#1){\ifnum\count1=0 {\bf #1}\else \ref{#1}\fi}
\def\bea(#1){\ifnum \count1=0   $$ \begin{array}{#1}
\else \begin{equation} \begin{array}{#1} \fi}
\def\eea{\ifnum \count1=0 \end{array} $$
\else  \end{array}\end{equation}\fi}
\def\elea(#1){\ifnum \count1=0 \end{array}\label{#1}\eqno({\bf #1}) $$
\else\end{array}\label{#1}\end{equation}\fi}
\def\cit(#1){
\ifnum\count1=0 {\bf #1} \cite{#1} \else
\cite{#1}\fi}
\def\bibit(#1){\ifnum\count1=0 \bibitem{#1} [#1    ] \else \bibitem{#1}\fi}
\def\ds{\displaystyle}
\def\hb{\hfill\break}
\def\comment#1{\hb {***** {\em #1} *****}\hb }

\newcommand{\TZ}{\hbox{T\hspace{-5pt}T}}
\newcommand{\MZ}{\hbox{I\hspace{-2pt}M}}
\newcommand{\ZZ}{\hbox{Z\hspace{-3pt}Z}}
\newcommand{\NZ}{\hbox{I\hspace{-2pt}N}}
\newcommand{\RZ}{\hbox{I\hspace{-2pt}R}}
\newcommand{\CZ}{\,\hbox{I\hspace{-6pt}C}}
\newcommand{\PZ}{\hbox{I\hspace{-2pt}P}}
\newcommand{\QZ}{\hbox{I\hspace{-6pt}Q}}
\newcommand{\HZ}{\hbox{I\hspace{-2pt}H}}
\newcommand{\EZ}{\hbox{I\hspace{-2pt}E}}
\newcommand{\GZ}{\,\hbox{l\hspace{-5pt}G}}

\vbox{\vspace{38mm}}
\begin{center}
{\LARGE \bf Mirror Symmetry of elliptic curves and Ising
Model }\\[5mm]

Shi-shyr Roan
\footnote{Supported in part by the NSC grant of Taiwan.}\\{\it Institute of
Mathematics \\ Academia Sinica \\
Taipei , Taiwan \\ (e-mail: maroan@ccvax.sinica.edu.tw)} \\[5mm]
\end{center}

\begin{abstract} We study the differential equations
governing mirror symmetry of elliptic curves, and obtain
a characterization of the ODEs which give rise to the integral
${\bf q}$-expansion of mirror maps. Through theta function
representation of the defining equation, we express the mirror
correspondence in terms of theta constants. By investigating
the elliptic curves in $X_9$-family, the identification of
the Landau-Ginzburg potential with the spectral curve of
Ising model is obtained. Through the Jacobi
elliptic function parametrization of Boltzmann
weights in the statistical model,
an exact Jacobi form-like formula of mirror map
is described .
\end{abstract}

\vfill
\eject

\section{Introduction}
Recent progrss in physcists' construction \cite{COGP} \cite{Y}
of the "number" of
rational curves of an arbitrary degree on a large class of
Calabi-Yau spaces has stimulated efforts to find a mathematical
understanding of this remarkable "counting" principle. As is
known the main ingredient of practically all examples is to
express the "counting" function, called the mirror map, in
terms of solutions of a generalized hypergeometric system.
As a physical theory, it is the N=2 supersymmetry (SUSY) two-dimensional
Landau-Ginzburg (LG) models to describe the mirror symmetry
of $\sigma$-models on K\"{a}hler manifolds with vanishing
first Chern class , ( for the basic notion of
mirror symmetry, we refer readers to \cite{Y}). This novel
principle gives also counting functions on other $c_1$=0 algebraic manifolds of
an arbitrary (complex) dimension. In the elliptic curve case,
there are three ways of realizing them as hypersurfaces in weighted
projective 2-space, and the moduli parameter is always
connected to the classical $J$-function by an algebraic relation
\cite{KLRY} \cite{LLW}:
$$
\put(-220, 20){\line(0, -1){90}}
\put(200, 20){\line(0, -1){90}}
\put(-220, 20){\line(1, 0){420}}
\put(-220, 0){\line(1, 0){420}}
\put(-195, 20){\line(0, -1){90}}
\put(-185, 5){ \shortstack{ Constraint}}
\put(15, 5){ \shortstack{ Differential \ operator }}
\put(150, 5){ \shortstack{
$1728 J( z )$ }}
\put(-215, -20){\shortstack{$P_8$}}
\put(-195, -20){ \shortstack{ $x_1^3 + x_2^3 + x_3^3 - z^{-1/3}
x_1x_2x_3 = 0$ in $\PZ^2_{(1,1,1)}$ }}
\put(15, -20){ \shortstack{ $\Theta^2 -
3 ( 3 \Theta + 2 ) ( 3 \Theta + 1 )$ }}
\put(150, -20){ \shortstack{
$\frac{( 1 + 216z)^3}{ z(1-27z)^3 }$ }}
\put(-215, -40){\shortstack{$X_9$}}
\put(-195, -40){ \shortstack{ $x_1^4 + x_2^4 + x_3^2 - z^{-1/4}
x_1x_2x_3 = 0$ in $\PZ^2_{(1,1,2)}$ }}
\put(15, -40){ \shortstack{ $\Theta^2 -
4 ( 4 \Theta + 3 ) ( 4 \Theta + 1 )$ }}
\put(150, -40){ \shortstack{
$\frac{( 1 + 192z)^3}{ z(1-64z)^2 }$ }}
\put(-215, -60){\shortstack{$J_{10}$}}
\put(-195, -60){ \shortstack{ $x_1^6 + x_2^3 + x_3^2 - z^{-1/6}
x_1x_2x_3 = 0$ in $\PZ^2_{(1,2,3)}$ }}
\put(15, -60){ \shortstack{ $\Theta^2 -
12 ( 6 \Theta + 5 ) ( 6 \Theta + 1 )$ }}
\put(150, -60){ \shortstack{
$\frac{1}{ z(1-432z) }$ }}
\put(-220, -70){\line(1, 0){420}}
\put(-35, -90){ \shortstack{ Table (I)}}
$$
Here the differential operator describes the Picard-Fuchs equation
for the family, and $\Theta : = z \frac{ \partial}{
\partial z }$. With the variable ${\bf t}$ obtained by a ratio of fundamental
solutions
of the differential equation near $z = 0$, the mirror map
yields the following numerical expansion of ${\bf q}
  \ ( : = e^{2\pi i {\bf t}}  \ ) $ for the parameter $z$ :
\bea(cl)
P_8  : z ( {\bf q} ) = & {\bf q} - 15 {\bf q}^2 +
171 {\bf q}^3 - 1679 {\bf q}^4 + 15054 {\bf q}^5
- 126981 {\bf q}^6 + \ldots \\
X_9  : z ({\bf q }) = & {\bf q} - 40 {\bf q}^2 +
1324 {\bf q}^3 - 39872 {\bf q}^4 + 1136334 {\bf q}^5
- 31239904 {\bf q}^6 + \ldots \\
J_{10}  : z ( {\bf q }) = & {\bf q} - 312{\bf q}^2 +
87084 {\bf q}^3 - 23067968 {\bf q}^4 +
5930898126 {\bf q}^5 - 1495818530208 {\bf q}^6 + \ldots
\elea(qexp)
Note that the "counting" numbers in these expansions are all integers.
For a general Calabi-Yau hypersurface family in a weighted projective
space, one also produces a "counting" function of such kind.
However the mathematical reason for the arithmetical
nature of "counting" functions is poorly understood, but
a fundamental understanding of the counting principle should
be important to further mathematical
development of mirror symmetry.
In \cite{KLRY}, the generalized Schwarzian equations were
derived for mirror maps
of one-modulus cases as one effort towards this direction.
The starting point
of the present work is to clarify the role of
differential equations in the integral property of the
counting function $z ( {\bf q} )$. We find a charactrization
of the equations appeared in Table (I) by their qualitative
relations with the $J$-function. For the precise statement of
the result, see Theorem 1 of the context.
On the other hand, the numerical evidence
has also suggested $z ({\bf q} )$ might possess
a certain structure as modular functions. To the author's
knowledge, not much is known
about the exact modular form-like expression of
$z ( {\bf q} )$, even on elliptic curve cases. In this paper,
we have obtained the elliptic theta function parametrization
of the constraint, i.e. LG superpotential, in Table (I) , and
also the exact formula of $z ( {\bf q} )$ in terms of theta
constants.
The key ingredient is the observation of the connection
between discrete symmetries encoded in the constraint,
 and their hidden theta function ( projective )
representations. Our purpose here is
to extensively analyse the discrete symmetries appeared
in Table (I),
and to determine the theta function parametrization of the
superpotential for each case, which allows one to obtain the
exact formula of the moduli parameter.
One main contribution of the present work is that we have
connected $X_9$-family with Ising model,
 a standard physical theory which has been served as a basis to
provide a simple 2-dimensional statistical model.
Here the Jacobi elliptic parametrzation of Boltzmann weights in
Ising model is used for the derivation of theta function
representation of the $X_9$-potential, and the Jacobi form
expression of temperature-like parameter of Ising family leads
to a closed form of $z({\bf q})$ for $X_9$ in (\req(qexp)) in
terms of theta constants. With this novel phenomena,
it becomes increasingly interesting in the interplay of geometry of
$c_1$=0 K\"{a}hler manifolds and other 2-dimensional solvable
statistical models. As it is well known, theta function
parametrizations have provided a powerful tool in 2-dimensional
lattice models to obtain quantities of physical interest
\cite{B} \cite{TF79}. In recent years, there has been
considerable progress in the study of chiral Potts $N$-state
models \cite{AMP} \cite{BBP}, as a generalization of
Ising model. The Boltzmann weights of the chiral Potts models lie
on hyperelliptic curves with a large number of discrete symmetries, and
their theta function parametrizations are known in
\cite{Bax} \cite{R92}. The question that we address for
future investigation is to establish a connection
between this hyperellitic function parametrization with mirror map of
Calabi-Yau spaces. A resoluton might point towards some
future structure, yet to be explained.

The following is a summary of the contents of this article:
In Sect. 2, we recall some basic facts on elliptic theta
functions and Heisenberg group representation, which will be needed
for the discussion of this paper.
In Sect. 3, we study the Schwarzian equations satisfied by
the mirror map, which are derived from a special
type of Fuchsian differential equations
\cite{KLRY} \cite{LY}. We
characterize the differential operators in Table (I), which are
solely governed by the integral property of the ${\bf q}$-expansion
of the Schwarz triangle function, and its qualitative
relation with $J$-function. Also we indicate the
$J_{10}$-family as an equivalent version of Weierstrass form
of elliptic curves.
In Sect. 4, the elliptic theta function
parametrization of $P_8$-family is derived, so is the expression of
$z ( {\bf q} )$ in terms of theta constants.
Based on the identification of symmetries of the defining
equation with the finite Heisenberg group of degree 3, the standard theta
function representation of the group gives rise to the
parametrization of $P_8$-potential. In Sect. 5, we give a
brief review on elliptic curve theory related to Boltzamann
weights of Ising model, which will be relevant to our
discussion. Primary focus
is on its Jacobi elliptic function parametrization.
With this parametrization, by examining the relation
between $X_9$-potential with
Ising model  we derive the
Jacobi elliptic function representation of elliptic curves
in $X_9$-family in Sect. 6, and also the exact formula for the moduli
parameter $z ( { \bf q} )$. After carrying out the mathematical
results of this paper, finally in Sect. 7  we will mention a
comparison of some essential structures in two physical theories: N=2 SUSY LG
theory and
exactly solvable statistical model, the geometry of which is
respectively presented in Calabi-Yau spaces and hyperelliptic
curves of chiral Potts models.

{\bf Notations } \par \noindent
$\HZ = \{ \tau \in \CZ \ | \ \mbox{Im} ( \tau ) > 0  \} \
$ \  the complex upper-half plane, which is acted by
$SL_2 ( \RZ )$ via fractional transformations.
\par \noindent
$ \Gamma  = SL_2 ( \ZZ ) \ $ .
\par \noindent
$\Gamma ( m )  =  \{ \gamma \in \Gamma \ | \ \gamma \equiv 1_2
\ ( \mbox{mod \ } m ) \} $ the  principal
congruence  subgroup  of level $m, \ m \in
\ZZ_{ + } $. \par \noindent
$\EZ_{a,b}$ = the 1-dimensional torus $\CZ / ( \ZZ a + \ZZ b )$
for two $\RZ$-independent complex numbers $a , b$.
\par \noindent
$\EZ_{a,b} ( d )$ = the $d$-torsion of $\EZ_{a,b}$ for a positive
integer $d$.

\section{ Preliminary }
Here we recall the definitions of Heisenberg group and
theta functions, and list some of their basic properties
that will be used in the context of this paper. For the
details , we refer the readers to some standard text books on
theta functions,  e.g. \cite{M}. \par \noindent
{\bf Definition . }

(i). $\GZ = {\CZ}^*_1 \times {\RZ} \times {\RZ} \ , \
 ( {\CZ}^*_1 = \{ \alpha \in {\CZ}^* \ | \ | \alpha | = 1 \} )
$ : the ( 3-dimensional ) Heisenberg group
with the group law :
\[
( \alpha , \delta , \mu ) \cdot ( \alpha' , \delta' , \mu' )
= ( \alpha \alpha' e^{ 2 \pi i \mu \delta'} , \delta +
\delta' , \mu + \mu' ) \ .
\]

(ii). For $p , q \in {\QZ} - \{ 0 \}$,
$\Lambda ( p , q )$ = the subgroup of $ \GZ$ generated
 by $( 1 , p , 0 ) , ( 1 , 0 , q )$.

(iii). $\GZ_d = {\nu}_d \times {\ZZ}_d \times {\ZZ}_d \ , \
 ( \ {\nu}_d = \{ \alpha \in {\CZ}^* \ | \  \alpha^d = 1 \} )
$ : the finite Heisenberg group of degree $d$
with the group law
\[
( \alpha , \overline{ \delta } , \overline{ \mu } ) \cdot
( \alpha' , \overline{ \delta' } , \overline{ \mu' } )
= ( \alpha \alpha' e^{ 2 \pi i \mu \delta'} , \overline{ \delta +
\delta' } , \overline{ \mu + \mu' } ) .
\]

(iv). $\widetilde{\GZ}_d = \GZ_d \cdot \ZZ_2
$ : the extended degree $d$ Heisenberg group which is the
semidirect product of $\GZ_d$ and $\ZZ_2$ with
$\GZ_d $ normal in $ \widetilde{\GZ}_d$,
where the conjugate action on $\GZ_d$ of
the non-trivial element of $\ZZ_2$ is given by
\[
( \alpha ,
\overline{ \delta } , \overline{ \mu } ) \mapsto
( \alpha , - \overline{ \delta } , - \overline{ \mu } ) \ , \
\mbox{for \ }
( \alpha , \overline{ \delta } , \overline{ \mu } ) \in
\GZ_d \ .
\]

(v). The canonical representaton of $\widetilde{\GZ}_d$ is by
definition the $d$-dimensional irreducible representation of
$\widetilde{\GZ}_d$ where group elements act on a basis
$\{ e_k \}_{k=0}^{d-1}$ of the vector space by
\bea(llcrl)
(( \alpha , 0 , 0 ) \times 0)  e_k & = & \alpha
e_k , & & \\
(( 1 , \overline{1} , 0 ) \times 0) e_k & = &
e_{k+1} , &
(( 1 , 0 , \overline{1} ) \times 0) e_k  & =
 e^{ 2 \pi i \frac{k}{ d } } e_k , \\
(( 1 , 0 , 0 ) \times \overline{1} ) e_k & = &
e_{d-k} \ ,&   \mbox{for \ } 0 \leq k \leq d-1 , &
 e_d : = e_0 \ .
\elea(canrep)
$\Box$ \par \vspace{.2in} \noindent
One has the exact sequence of groups:
\be
1 \longrightarrow \Lambda ( 1 , d ) \longrightarrow \Lambda
( \frac{1}{d} , 1 ) \stackrel{ \phi }{ \longrightarrow} \GZ_d
\longrightarrow 1 \ ,
\ele(LGd)
where $\phi$ is the group homomorphism with
\[
\phi ( 1 , \frac{1}{d} , 0 ) = (1 , \bar{1} , 0 ) , \ \
\phi ( 1 , 0 , 1 ) = (1 , 0 , \bar{1}  ) .
\]
For $\delta, \mu \in \RZ , \tau \in \HZ$ and an entire function
$f$ on $\CZ $, one defines the functions $ S_{\mu}f$ and
$T_{\delta} ( \tau ) f $ by
$$
\begin{array}{clll}
 ( S_{\mu}f )  ({\rm z}) &= &f ( {\rm z} + \mu ) ,& \\
 ( T_{\delta} ( \tau ) f ) ( {\rm z} ) &= &
q^{\delta^2}  e^{ 2 \pi i \delta {\rm z} }
f ( {\rm z} + \delta \tau ), & \mbox{for} \ {\rm z} \in \CZ  .
\end{array}
$$
Then $S_{\mu}$ and $T_{\delta}( \tau )$ acts on the space of
entire functions with the relations,
\be
S_{\mu}  S_{\mu'} = S_{\mu + \mu' } \ , \ \
T_{\delta} ( \tau ) T_{\delta'} ( \tau )  =
 T_{\delta + \delta' } ( \tau )  \ , \ \
S_{\mu} T_{\delta} ( \tau ) =
e^{ 2 \pi i  \mu \delta } T_{\delta} ( \tau ) S_{\mu}  ,
\ele(STTS)
and they generate
a representation of the Heisenberg group $\GZ$ by
\[
( \alpha , \delta , \mu ) \circ f  : = \alpha  T_{\delta}
(\tau) S_{\mu} f  \ , \
\mbox{for} \ ( \alpha , \delta , \mu ) \in \GZ\ , \
f: \CZ \rightarrow \CZ \ .
\]
We shall also write $T_{ \delta }$ instead of
$T_{ \delta }( \tau )$ when no confusion arises.
For $\tau \in \HZ $, the the theta function
$\vartheta ( {\rm z} , \tau )$ of $\EZ_{\tau , 1 }$ and the
theta function $\vartheta [ \begin{array}{c} \delta \\ \mu
\end{array} ] ( {\rm z} , \tau )$ with characteristics $
\delta , \mu \in {\RZ}$ are defined by
$$
\begin{array}{clll}
\vartheta ( {\rm z} ) \ \ ( = \vartheta ( {\rm z} , \tau ) ) & : =
\sum_{m \in {\ZZ}} ( T_n  1 ) ( {\rm z} ) &
= \sum_{m= \infty }^{
\infty } q^{m^2} e^{ 2 \pi i m {\rm z} } \ ,&  \\
\vartheta [ \begin{array}{c} \delta \\ \mu
\end{array} ] ( {\rm z} , \tau ) & : = S_{ \mu } T_{ \delta }
\vartheta ( {\rm z} )  &=
q^{ \delta^2} e^{ 2 \pi i \delta ( {\rm z} + \mu ) }
\vartheta( {\rm z} + \delta \tau + \mu )  ,& {\rm z} \in {\CZ} ,
\end{array}
$$
where $1$ is the function with constant
value one . Then $\vartheta ( {\rm z} , \tau) $ is the unique
entire function invariant under $\Lambda ( 1 , 1 )$ , and
we have
\be
 T_{ \alpha }  \vartheta [ \begin{array}{c} \delta \\ \mu
\end{array} ]  = e^{ - 2 \pi i \alpha \mu } \vartheta [ \begin{array}{c} \delta
+ \alpha \\ \mu
\end{array} ]  \ , \ \ S_{ \beta } \vartheta [
\begin{array}{c} \delta \\ \mu  \end{array} ]  =
\vartheta [ \begin{array}{c} \delta \\ \mu + \beta
\end{array} ]  .
\ele(TSThe)
It is known that the theta function has the following
representation of infinite product:
\[
\vartheta ( {\rm z} , \tau )   =   \prod_{n=1}^{ \infty} (
1 + 2 q^{2n-1} \cos 2 \pi {\rm z} + q^{4n-2} ) ( 1 - q^{2n} ) \ , \ \
q := e^{ \pi i \tau } \ ,
\]
and satisfies the quasi-periodicity and evenness
relations:
$$
\begin{array}{ccl}
\vartheta ( {\rm z} + 1 ) & = & \vartheta ( {\rm z} ) , \\
\vartheta ( {\rm z} + \tau ) & = &
q^{-1}e^{-2 \pi i {\rm z}} \vartheta ( {\rm z} ) , \\
\vartheta ( - {\rm z} ) & = & \vartheta ( {\rm z} ) .
\end{array}
$$
Note that the variable ${\bf q}$ in Sect. 1 relates to the
above $q$ by
\[
{\bf q} = q^2 \ .
\]
The following relations hold for
$\vartheta [ \begin{array}{c} \delta \\ \mu
\end{array} ]$:
\bea(lcll)
\vartheta [ \begin{array}{c} \delta \\ \mu
\end{array} ] ( {\rm z} + 1 , \tau ) & = & e^{ 2 \pi i  \delta}
\vartheta [ \begin{array}{c} \delta \\ \mu
\end{array} ] ( {\rm z} , \tau ) , & \\
\vartheta [ \begin{array}{c} \delta \\ \mu
\end{array} ] ( {\rm z} +  \tau , \tau ) & = & q^{-1} e^{ - 2
\pi i  ( {\rm z} +  \mu ) }
\vartheta [ \begin{array}{c} \delta \\ \mu
\end{array} ] ( {\rm z} , \tau ) , &  \\
\vartheta [ \begin{array}{c} \delta \\ \mu
\end{array} ] ( {\rm z}  , \tau ) = 0 & \Longleftrightarrow &
{\rm z} \equiv ( \frac{1}{2} - \delta ) \tau + ( \frac{1}{2} - \mu ) \
\ \ &( \mbox{mod} \ \ {\ZZ} \tau + {\ZZ} ) \ .
\elea(prthetacha)
We have
\bea(lll)
\vartheta [ \begin{array}{c} \delta' + \delta \\ \mu' + \mu
\end{array} ] ( {\rm z} , \tau ) & = & q^{ \delta^2}
e^{ 2 \pi i \delta ( {\rm z} + \mu + \mu' )}
\vartheta [ \begin{array}{c} \delta' \\ \mu'
\end{array} ] ( {\rm z} + \delta \tau + \mu , \tau ) \\
\vartheta [ \begin{array}{c} \delta + 1 \\  \mu
\end{array} ] ( {\rm z} , \tau ) & = &
\vartheta [ \begin{array}{c} \delta \\ \mu
\end{array} ] ( {\rm z}  , \tau ) , \\
\vartheta [ \begin{array}{c}  \delta \\  \mu + 1
\end{array} ] ( {\rm z} , \tau ) & = & e^{ 2 \pi i \delta }
\vartheta [ \begin{array}{c} \delta \\ \mu
\end{array} ] ( {\rm z} , \tau ) .
\elea(relthetach)
Hence
\bea(llll)
\vartheta [ \begin{array}{c} \delta \\ \mu
\end{array} ] ( - {\rm z}  , \tau ) & = &
\vartheta [ \begin{array}{c} 1 - \delta \\ - \mu
\end{array} ] ( {\rm z}  , \tau ) &
= e^{ - 2 \pi i \delta } \vartheta [ \begin{array}{c} 1 - \delta \\ 1 - \mu
\end{array} ] ( {\rm z} , \tau )   \\
\vartheta [ \begin{array}{c} 1 - \delta \\ 1 - \mu
\end{array} ] ( \frac{ \tau + 1 }{2} , \tau )  & = &
- e^{ 2 \pi i  \mu  }
\vartheta [ \begin{array}{c} \delta \\ \mu
\end{array} ] ( \frac{ \tau + 1 }{2}  , \tau ) &.
\elea(-zzero)
The infinite product representation of four theta functions
with half-integer characteristics are given by
\bea(lllll)
 \vartheta_1  ( {\rm z}, \tau )  & : = &
\vartheta [ \begin{array}{c} \frac{1}{2} \\ \frac{1}{2}
\end{array} ] ( {\rm z} , \tau ) & = &
2 q_0 q^{ \frac{1}{4}} \sin \pi {\rm z} \prod_{n=1}^{ \infty} (
1 - 2 q^{2n} \cos 2 \pi {\rm z} + q^{4n} )  , \\
 \vartheta_2 ( {\rm z}  , \tau )  & : = &
\vartheta [ \begin{array}{c} \frac{1}{2} \\ 0
\end{array} ] ( {\rm z} , \tau ) & = &
2 q_0 q^{ \frac{1}{4}} \cos \pi {\rm z} \prod_{n=1}^{ \infty} (
1 + 2 q^{2n} \cos 2 \pi {\rm z} + q^{4n} )  ,
 \\
 \vartheta_3 ( {\rm z}  , \tau )  & : = &
\vartheta [ \begin{array}{c} 0 \\ 0
\end{array} ] ( {\rm z} , \tau ) & = &
q_0 \prod_{n=1}^{ \infty} ( 1 + 2 q^{2n-1} \cos 2 \pi {\rm z} +
q^{4n-2} ) ,
\\
 \vartheta_4 ( {\rm z}  , \tau )  & : = &
\vartheta [ \begin{array}{c} 0 \\ \frac{1}{2}
\end{array} ] ( {\rm z}, \tau ) & = & q_0  \prod_{n=1}^{ \infty} (
1 - 2 q^{2n - 1} \cos 2 \pi {\rm z} + q^{4n - 2} ) ,
\elea(prodformtheta)
where $ q_0 : = \prod_{n=1}^{\infty} ( 1 - q^{2n} ) $.
The functions $\vartheta_1 , \vartheta_2 ,
\vartheta_3 , \vartheta_4 $ have zeros at $ {\rm z} = m + n \tau ,
m + \frac{1}{2} + n \tau , m + \frac{1}{2} +
(n + \frac{1}{2}) \tau , m  +
(n + \frac{1}{2}) \tau $ respectively for integers $m$ and $n$
, with $\vartheta_1 $
odd function and $\vartheta_2 , \vartheta_3 ,
\vartheta_4 $ even functions of the variable z.
The above theta functions satisfy the square relations:
\bea(ll)
\vartheta_1^2 ( {\rm z} , \tau ) \vartheta_2^2 ( 0 , \tau) =
& \vartheta_4^2 ( {\rm z} , \tau )
\vartheta_3^2 ( 0 , \tau) - \vartheta_3^2 ( {\rm z} , \tau )
\vartheta_4^2 ( 0, \tau ) , \\
\vartheta_1^2 ( {\rm z}, \tau ) \vartheta_3^2 ( 0, \tau ) =
& \vartheta_4^2 ( {\rm z}, \tau )
\vartheta_2^2 ( 0, \tau ) - \vartheta_2^2 ( {\rm z}, \tau )
\vartheta_4^2 ( 0, \tau ) , \\
\vartheta_1^2 ( {\rm z}, \tau ) \vartheta_4^2 ( 0, \tau ) =
& \vartheta_3^2 ( {\rm z}, \tau )
\vartheta_2^2 ( 0, \tau ) - \vartheta_2^2 ( {\rm z}, \tau )
\vartheta_3^2 ( 0, \tau ) , \\
\vartheta_4^2 ( {\rm z}, \tau ) \vartheta_4^2 ( 0, \tau ) =
& \vartheta_3^2 ( {\rm z}, \tau )
\vartheta_3^2 ( 0, \tau ) -
\vartheta_2^2 ( {\rm z}, \tau ) \vartheta_2^2 ( 0, \tau ) ,
\elea(relth)
hence the identity of the theta functions
of zero argument:
\be
  \vartheta_2^4 ( 0 , \tau ) + \vartheta_4^4 ( 0 , \tau ) =
\vartheta_3^4 ( 0 , \tau )  .
\ele(0THe)
For a positive integer $d$, let $\mbox{Th}_d ( \tau )$ be the
space of theta functions with characteristics which
are invariant under the group $\Lambda ( 1 , d )$. Then
$\mbox{Th}_d ( \tau )$ is a $d$-dimensional vector space with the basis
\[
\vartheta [ \begin{array}{c} \frac{k}{d} \\ 0
\end{array} ] = \vartheta [ \begin{array}{c} \frac{k}{d} \\ 0
\end{array} ] ( {\rm z} , \tau ) \ , \ k = 0 , \ldots , d-1 .
\]
By (\req(TSThe)) (\req(relthetach)), one has
\[
T_{\frac{1}{d}} \vartheta [ \begin{array}{c} \frac{k}{d} \\ 0
\end{array} ] = \vartheta [ \begin{array}{c} \frac{k+1}{d} \\ 0
\end{array} ] \ , \ \
S_1 \vartheta [ \begin{array}{c} \frac{k}{d} \\ 0
\end{array} ] = e^{ \frac{2 k \pi i}{ d } }
\vartheta [ \begin{array}{c} \frac{k}{d} \\ 0
\end{array} ] \ .
\]
By (\req(canrep)) (\req(LGd)) and (\req(-zzero)) , the action
of $T_{\frac{1}{d}}$ and $S_1$ on $\mbox{Th}_d ( \tau )$ , together
with the involution,
\[
\vartheta [ \begin{array}{c} \frac{k}{d} \\ 0
\end{array} ] ( {\rm z} ) \mapsto
\vartheta [ \begin{array}{c} \frac{k}{d} \\ 0
\end{array} ] ( - {\rm z} ) \ ,
\]
gives rise to the canonical
representation of the extended degree $d$ Heisenberg group
$\widetilde{\GZ}_d$ via the identification:
\[
e_k = \vartheta [ \begin{array}{c} \frac{k}{d} \\ 0
\end{array} ] \ \mbox{ for \ } 0 \leq k \leq d \ .
\]

\section{ ODEs for Mirror Map of Elliptic Curves}
In this section, we shall characterize the differential operators
in Table (I) through $J$-function with an emphasis on
the property of integral ${\bf q}$-expansion of
the variable $z$ . It is known that an elliptic curve can be
represented in the Weierstrass form:
\be
y^2 = 4 x^3 - g_2 x - g_3 \ , \ ( x , y ) \in \CZ^2 \ ,
\ele(Weierf)
with the parameter
\[
[ g_2 , g_3 ] \in \PZ_{(2, 3)}^1 \ ,
\]
equivalently the value
\[
J := \frac{ g_2^3 }{ g_2^3 - 27 g_3^2 }
\]
in $\CZ \cup \{ \infty \}$, which is isomorphic to
the Riemann surface $\Gamma \backslash \HZ$ via the modular function
$J ( \tau )$ of level 1. The periods of
elliptic curves satisfy the Picard-Fuchs
equation:
\[
\frac{ d^2 y }{ dJ^2} + \frac{1}{J} \frac{ d y }{ dJ }
+ \frac{31J - 4}{144J^2 ( 1 - J )^2} y = 0 \ ,
\]
hence are expressed by the Riemann $P$-function:
\[
P \left\{ \begin{array}{cccr}
0 & 1 & \infty & \\
\frac{1}{6} & \frac{1}{4} & 0 & \ ; J \\
\frac{-1}{6} & \frac{3}{4} & 0 &
\end{array}
\right\} \ .
\]
The ratio of two periods gives the variable $\tau$ of $\HZ$,
which as a function of $J$, satisfies the well-known Schwarzian
equation:
\[
\{ \tau , J  \} = \frac{3}{8 ( 1 - J )} +
\frac{4}{9J^2} + \frac{23}{72 J(1-J)} \ .
\]
Here the Schwarzian differential operator on the
left hand side is defined by:
\[
  \{ y , x \} = \frac{ y'''}{ y'} - \frac{3}{2} (
\frac{y'' }{y'} )^2 \ .
\]
The inverse function $J( \tau )$ of $\tau ( J )$ is a modular function
\cite{G} \cite{Sh}, expressed by :
\be
1728 J( \tau )  =  \frac{ g_2 ( \tau )^3}
{ g_2 ( \tau )^3 - 27 g_3 ( \tau )^2 }
\ ,  \ \ \mbox{where} \  \
g_2 ( \tau  ) = 60 G_4 ( \tau ) \ , \
\ g_3 ( \tau ) = 140 G_6 ( \tau ) \ ,
\ele(Jg23)
and $G_{2k}$'s are the Eisenstein series with the Fourier
expansion:
\[
G_{2k} ( \tau ) = \frac{2 (-1)^k ( 2 \pi )^{2k} }{ ( 2k - 1 ) !}
\{ \frac{(-1)^k B_{2k} }{4k} + \sum_{n=1}^{\infty} n^{2k-1}
\frac{{\bf q}^n}{1 - {\bf q}^n} \} \ , \ \ B_{2k} :
\mbox{ the \ Bernoulli \ number } \ .
\]
Hence $1728J( \tau )$
admits an integral ${\bf q}$-series :
\[
1728 J( \tau ) = {\bf q}^{-1} + 744 + 196884 {\bf q} +
21493760 {\bf q}^2 + \cdots   \ \ \ ,
\ \
{\bf q} = e^{ 2 \pi i \tau } .
\]
In this section, we shall discuss the integral property of the
${\bf q}$-series in (\req(qexp)) and
the relation between $J$ and $z$ in Table (I). We state
the following simple lemma for later use.
\par \vspace{0.2in} \noindent
{\bf Lemma 1. } Let
\[
w ( z ) = z + \sum_{m \geq 2} a_m z^m \ , \ \
z ( {\bf q} ) = {\bf q} + \sum_{m \geq 2} k_m {\bf q}^m \
\]
be formal power series with $a_m \in \ZZ$ for all $m \geq 2$.
Then  $w ( z ( {\bf q} ) )$
has an integral ${\bf q}$-expansion if and only if
all the $k_m$'s are integers.
$\Box$ \par \vspace{0.2in} \noindent

Consider the following ordinary differential equations of
Fuchsian type:
\be
( \Theta^2 - \lambda z ( \Theta + \alpha ) ( \Theta + \beta ))
y ( z ) = 0 \ ,
\ele(eqz)
where $\Theta = z \frac{d}{dz}$ and $\lambda, \alpha , \beta$
are positive rational numbers with
\[
\alpha + \beta = 1 \ , \ \ \alpha > \beta \ .
\]
The equation (\req(eqz)) is invariant under the change of
variables
\[
z \mapsto - z + \frac{1}{\lambda} \ ,
\]
with three regular singular points,
\[
z = 0 , \frac{1}{\lambda } , \infty \ .
\]
Its solutions are expressed by the Riemann $P$-function:
\[
P \left\{ \begin{array}{cccr}
0 & \frac{1}{\lambda } & \infty & \\
0 & 0 & \alpha & \ ; z \\
0 & 0 & \beta &
\end{array}
\right\} \ .
\]
By the change of variables, $x = \lambda z$, (\req(eqz))
becomes the hypergeometric equation:
\be
x ( 1- x ) \frac{d^2 y}{ dx^2} + ( 1 - 2 x ) \frac{d y}{ dx}
- \alpha \beta y = 0 \ ,
\ele(hyeq)
whose fundamental solutions at $x = 0$ are given by the
hypergeomertic series
\[
y_1  ( x ) = F ( \alpha, \beta; 1; x ) \ ,
\]
together with another solution, uniquely determined by the
form
\[
y_2 ( x ) = \log ( x ) F ( \alpha, \beta; 1; x ) +
\sum_{n = 1}^{\infty} a_n x^n \ .
\]
The local system for the equation (\req(hyeq))
is described
by analytic continuations of $y_1 ( x )$ and
$y_2 ( x )$, or of any other fundamental solutions:
\[
a y_1 ( x ) + b y_2 ( x ) \ \ \mbox{and} \ \ c y_1 ( x ) + d y_2 ( x ) \ ,
\ \ \left( \begin{array}{cc}
a&b\\
c&d
\end{array}
\right) \in SL_2 ( \CZ ) \ .
\]
The ratio
\[
t (x ) = \frac{a y_1 ( x ) + b y_2 ( x )}{a y_1 ( x ) +
b y_2 ( x )} \ ,
\]
is invariant under the substitution
\[
y ( x ) \mapsto  g ( x ) y ( x )
\]
for an arbitrary given function $g ( x )$, which transfers
the equation (\req(hyeq)) into another second order linear
differential equation. By choosing
\[
g ( x ) =  x^{ \frac{-1}{2}} ( 1 - x )^{ \frac{-1}{2}}
\ ,
\]
the equation is put into the form,
\[
 \frac{d^2 y }{dx^2} + Q ( x ) y = 0 \ , \ \ \mbox{with} \ \
Q ( x )  = \frac{ 1 - 4 \alpha \beta x ( 1 - x )}
{4 x^2 ( 1 - x )^2 } \ .
\]
Eliminating $y$ in the system of equations:
\[
\left\{ \begin{array}{l}
( \frac{d^2 }{dx^2} + Q ( x )) y = 0  \\
( \frac{d^2 }{dx^2} + Q ( x )) ( t y ) = 0 \ ,
\end{array}
\right.
\]
one obtains the non-linear Schwarzian differential equation for
$t ( x )$:
\be
\{ t , x \} = 2 Q ( x )  ,
\ele(Schtx)
whose solutions, known as Schwarz triangle functions,
are all equivalent under the action of $SL_2 (\CZ)$:
\[
t \mapsto \frac{a t + b }{ c t + d } \ , \
\left( \begin{array}{cc}
a&b\\
c&d
\end{array}
\right) \in SL_2 ( \CZ ) \ .
\]
Each solution gives rise to a local uniformization of the
punctured disc near $x = 0$. It determines the element
\[
t ( 0 ) : = \lim_{ x \rightarrow 0} t ( x )
\in \CZ \cup \{ \infty \} \ ,
\]
and a parabolic transformation fixing $t ( 0 )$, which is given by the local
monodromy around $x = 0$.
Therefore the solutions of equation (\req(eqz)) is expressed by
\[
P \left\{ \begin{array}{cccr}
0 & \frac{1}{ \lambda } & \infty & \\
0 & 0 & \alpha & \ ; z \\
0 & 0 & \beta &
\end{array}
\right\} = A f_1 ( z ) + B f_2 ( z ) \ , \ \ \ A , B \in \CZ ,
\]
with
$$
\begin{array}{lcl}
f_1 ( z ) = & y_1 ( \lambda z ) &= F ( \alpha, \beta; 1; \lambda z ) \ , \\
f_2 ( z ) = & y_2 ( \lambda z ) - \log ( \lambda )
y_1 ( \lambda z ) & = \log ( z ) f_1 ( z ) +
\sum_{n = 1}^{\infty} d_n z^n .
\end{array}
$$
The ratio
\be
{\bf t} ( z ) = \frac{ f_2 ( z ) }{ 2 \pi i f_1 ( z ) }
\ele(tz)
forms an uniformizing coordinate of the punctured disc at
$z = 0$, characterized as the solution of the equation
\be
\{ t , z \} = 2 \lambda^2 Q ( \lambda z ) \ ,
\ele(scheqn)
satisfying the conditions:
\be
 \lim_{ z \rightarrow 0} {\bf t} ( z ) = \infty \ , \ \
\lim_{ \theta \rightarrow 2 \pi^-} {\bf t} ( e^{i \theta } z ) =
{\bf t} ( z ) + 1 \ , \ \
\lim_{ z \rightarrow 0} \frac{e^{2 \pi i \bf t}}{ z } = 1 \ .
\ele(initial)
Denote
\[
{\bf q } = e^{ 2 \pi i {\bf t}} \ .
\]
We have an local isomorphism between the $z$-plane and
${\bf q}$-plane with the relation:
\be
z = {\bf q } + \sum_{ n \geq 2} k_n {\bf q }^n \ ,
\ k_n \in \CZ .
\ele(relzq)
Note that the coefficients $k_n$ depend on the data
$\lambda, \alpha$ analytically. The characterization of
the equations of type (\req(eqz)) with
integral coefficients for
its assocoated series (\req(relzq)), i.e.
\[
k_n \in \ZZ \  \ \ \forall \ n \ ,
\]
will be our main concern in what follows.

The analytical continuation of ${\bf t} ( z )$ gives rise to
a Riemann surface $\Re$,
which is an infinite cover over $z$-plane outside
$\{ 0 , \frac{1}{\lambda}, \infty \}$. By the non-zero Wronskian
determinant of the equation (\req(eqz)), one has
$\frac{ d {\bf t} }{ d z } \neq 0$. The projection of
$\Re$ to the ${\bf t}$-plane is a local isomorphism, hence
its image ${\bf t} ( \Re )$ forms a domain in $\PZ^1$. We
have the following relations between Riemann surfaces:
\bea(ccl)
\Re  & \stackrel{\bf t }{\longrightarrow} & {\bf t} ( \Re )
\subset \PZ^1 \\
z \downarrow \ \ \ & & \\
\PZ^1-\{ 0,\frac{1}{\lambda},\infty \}& &
\elea(Re)
One can extend $\Re$ to a Riemann surface over the zero-value
of $z$ as follows. Since the fundamental solutions of
the equation (\req(eqz)) near $z = \infty$ can take the form:
\[
z^{-\alpha}p_{\alpha} ( \frac{1}{z} ) \ , \ \
z^{-\beta}p_{\beta} ( \frac{1}{z} ) \ , \ \ |z| \gg 0 \ ,
\]
for $p_{\alpha}$ and $p_{\beta}$ power series in
$\frac{1}{z}$ with the constant term 1, on a connected region
of $\Re$ near $z = \infty$, one has
\be
{\bf t} ( z ) = \frac{ a z^{-\alpha}p_{\alpha} ( \frac{1}{z} )
+ b z^{-\beta}p_{\beta} ( \frac{1}{z} ) }
{c z^{-\alpha}p_{\alpha} ( \frac{1}{z} ) +
dz^{-\beta}p_{\beta} ( \frac{1}{z} )} \ , \ \ \
\left( \begin{array}{cc}
a&b \\ c&d
\end{array}
\right) \in SL_2 ( \CZ ) \ , \ \
\ \mbox{for} \
|z| \gg 0 \ .
\ele(tzexp)
Therefore
\[
\lim_{z \rightarrow \infty} {\bf t} ( z ) = \frac{b}{d} \in
\PZ^1 \ .
\]
Write
\[
  \alpha - \beta  = \frac{l}{k}
\]
with $k$ and $l$ two relatively prime positive integers, hence
$ k \geq 2$ .
By the expression of ${\bf t}$ in (\req(tzexp)),
there exist some local
coordinates $w$ near $z = \infty$, and $\tilde{t}$ near $
{\bf t}=\frac{b}{d}$ such that over a small punctured disc near
$z = \infty$, a connected region of $\Re$ is described by the
relation:
\[
w^l = \tilde{t}^k \ , \ \ ( w , \tilde{t} ) \neq ( 0 , 0 ) \ .
\]
Let $s$ be the local coordinate for the desingularization of the
above equation. The description of (\req(Re)) on a small
connected region over
a disc near $z = \infty$ is now equivalent to the following
diagram:
\bea(cclcl)
\{ 0 < | s | < \epsilon \}  & \longrightarrow &
\{ 0 < | \tilde{t} | < \delta \} & , & s \ \ \ \mapsto \ \tilde{t} = s^l \\
\downarrow  & & & & \downarrow \\
\{ 0 < | w | < \delta^\prime \} & & &,  & w = s^k
\elea(punswt)
hence we have
\be
\lim_{ \theta \rightarrow 2 \pi^{-} }
\tilde{t} ( z e^{i \theta } ) =
e^{2 \pi i ( \beta - \alpha )} \tilde{t} ( z ) \ .
\ele(z2pi)
Extending (\req(punswt)) to the following data,
\bea(cclcl)
\{ | s | < \epsilon \}  & \longrightarrow &
\{ | \tilde{t} | < \delta \} & , & s \ \ \ \mapsto \ \tilde{t} = s^l \\
\downarrow  & & & & \downarrow \\
\{ | w | < \delta^\prime \} & & &,  & w = s^k \ ,
\elea(swt)
one obtains a Riemann surface $\overline{\Re}$ as
a partial compactification
of $\Re$ with the extended diagram of (\req(Re)):
$$
\begin{array}{ccl}
\overline{\Re}  & \stackrel{\bf t }{\longrightarrow} &
{\bf t} ( \overline{\Re} ) \subset \PZ^1 \\
z \downarrow  & & \\
\PZ^1- \{ 0,\frac{1}{\lambda} \}& & \ .
\end{array}
$$
Note that the map $z$ branches at $z^{-1}( \infty )$ with
the multiplicity $k$.
On the other hand, the ( multi-valued ) function
$\tau ( J ) , J \neq 0 , 1 ,$
defines a Riemann surface $\Re_J$ with its partial
compactification $\overline{\Re_J}$ as follows:
$$
\begin{array}{ccl}
\overline{\Re_J}  & \stackrel{\tau }{\simeq} &
\HZ \subset {\PZ}^1 \\
J \downarrow \ \  & & \\
\CZ & & \ .
\end{array}
$$
We introduce a notion for the discussion of the relation
between ${\bf t}$ and ${\tau}$:
\par \vspace{0.2in} \noindent
{\bf Definition . } The Schwarz
triangle function ${\bf t}(z)$ (\req(tz)) for
the equation (\req(eqz)) is said to be related to
$J$-function  if for some morphisms $\Psi$ and $\Phi$,
the following diagram commutes:
\bea(rcccccl)
\PZ^1 & \supset & \PZ^1- \{ 0,\frac{1}{\lambda} \}&
\stackrel{ z }{\longleftarrow} & \overline{\Re}  & \stackrel{\bf t
}{\longrightarrow} &
{\bf t} ( \overline{\Re} )
\\
\Psi \downarrow \ \ & & \downarrow & & \Phi \downarrow \ \ &
& \ \parallel \\
\PZ^1 & \supset & \CZ & \stackrel{J }{\longleftarrow}
& \overline{\Re_J}  & \stackrel{\tau }{\simeq} &
\HZ \ \ \subset \ \PZ^1
\elea(compRRJ)
$\Box$ \par \vspace{0.2in} \noindent
{\bf Theorem 1. } All the differential operators of type
(\req(eqz)) whose ${\bf t}(z)$ is related to $J$-function
and with the integral ${\bf q}$-series $z ( {\bf q} )$ are
those listed in Table (I).
\par \vspace{0.2in} \noindent
The rest of this section will be mainly devoted to the proof of the
above theorem.
We shall regard a coordinate system of $\CZ$ as the (affine)
coordinate of the Riemann sphere $\PZ^1$ via the
identification:
\[
\PZ^1 = \CZ \cup \{ \infty \} \ .
\]
As before, by the change of variables $x = \lambda z$, the morphism
$\Psi$ in (\req(compRRJ)) induces the rational map
\[
\psi : \PZ^1 \longrightarrow \PZ^1 \ , \ \
x \mapsto J = \psi ( x ) \ .
\]
By examining the behavior over critical values of the function
$J ( \tau )$, the above morphism $\psi$ satisfies the following conditions:
\par \noindent
(i) The critical values of $\psi$ are contained in
$\{ 0 , 1 , \infty \}$. \par \noindent
(ii) $\psi^{-1} ( \infty ) = \{ 0 , 1 \}$, and
the multiplicity of $\psi$ at 0 is equal to 1, i.e.
\[
\mbox{mult}_{\psi} ( 0 ) = 1 \ .
\]
(iii) The value $\psi ( \infty )$ is equal to  0 or $1$.
For $x \neq \infty $ with $\psi ( x ) = 0 , 1 $, we have
\[
\mbox{mult}_{\psi} ( x ) =
\left\{ \begin{array}{ll}
3 & \mbox{if} \ \psi ( x ) = 0 , \\
2 & \mbox{if} \ \psi ( x ) = 1 .
\end{array}
\right.
\]
\par \vspace{0.2in} \noindent
{\bf Lemma 2 . } There are exactly 3 solutions for the above
$\psi$:
\[
\psi ( x ) =
\left\{ \begin{array}{l}
\frac{ ( 1 + 8x )^3}{ 64 x ( 1 - x )^3} \ , \\ [2mm]
\frac{ ( 1 + 3x )^3}{ 27 x ( 1 - x )^2} \ , \\ [2mm]
\frac{ 1}{ 4 x ( 1 - x )} \ .
\end{array}
\right.
\]
\par \vspace{0.2in} \noindent
{\it Proof . } Let $d$ be the degree of the map $\psi$.
By the conditions on $\psi$ , $d$ is greater than or equal to
2, and
\[
d = 2 \ \Longleftrightarrow \ \{ \mbox{critical \ value \ of \
} \psi \} = \{ 0 , 1 \} \ ,
\]
in which case, one has
$\psi ( \infty ) = 0$, hence easily see
\[
\psi ( x ) = \frac{ 1}{ 4 x ( 1 - x )}  \ .
\]
Now assume $ d \geq
3$. By Hurwitz Theorem, we have
\[
2d - 2 = r_0 + r_1 + r_{\infty} \ ,
\]
where $r_{j}$ is the sum of ramification indices of elements
in $\psi^{-1} (j)$. By (ii), $r_{\infty} = d-2$,
hence
\[
d = r_0 + r_1 \ .
\]
Let $k $ be the multiplicity of $\psi$ at $x = \infty  $. By (iii), we have
\[
( r_0 , r_1 ) =
\left\{ \begin{array}{ll}
( 2\frac{d-k}{3} + k - 1 \ , \ \frac{d}{2}) ,  &
\mbox{if} \ \psi ( \infty ) = 0 \ , \\ [2mm]
(2\frac{d}{3} \ , \ \frac{d-k}{2} + k - 1 ) ,  &
\mbox{if} \ \psi ( \infty ) = 1 .
\end{array}
\right.
\]
This implies that either
\[
d  = 4 \ , \ k = 1  \ , \ \ \ \psi ( x ) = \frac{a (x + b )^3}{x ( 1 - x )^3} \
,
\]
or
\[
d = 3 \ , \ k = 1 \ , \ \ \ \psi ( x ) = \frac{ (x + b )^3}{x ( 1 - x )^2} \ ,
\]
for some complex numbers $a \neq 0$, and $b \neq 0 , -1$. For
$d = 3$, there is only one critical point $x \in \PZ^1 - \{ 0 , 1,
\infty, -b \}$, which is given by $x = \frac{b}{3b+2}$. We have
\[
 \psi ( \frac{b}{3b+2} ) = 1 ,
\]
which implies
\[
0 = 27 b^3 + 27 b^2 - 4 = ( 3 b - 1 ) ( 3 b +2 )^2 \ , \ \
b = \frac{1}{3} .
\]
Therefore
\[
\psi ( x ) = \frac{ ( 1 + 3x )^3}{ 27x ( 1 - x )^2 }  \ .
\]
For $ d = 4 $ , there are exactly two critical points
$x_1, x_2$ not in $\{  0 , 1, \infty, -b \}$, and they satisfy
the following properties:
\[
\mbox{mult}_{\psi} ( x_i ) = 2 \ , \ \
\psi ( x_i ) = 1 \ \ \mbox{for \ } i = 1 , 2 .
\]
By the expression of $\psi ( x )$, one can easily see that
$x_i$'s are the solutions of
\[
x^2 + ( 2 + 4 b ) x - b = 0 ,
\]
whose discriminant is equal to
\[
16 b^2 + 20 b + 4 = 4 ( 4 b + 1 ) ( b + 1 ) \neq 0 \ .
\]
Then the following relations hold for
$x = x_1 , x_2$ :
$$
\begin{array}{l}
x + b = \frac{3 x ( 1 - x )}{ 1 - 4x } \ , \  \ \ \
x^3 = - b ( 2 + 4 b ) + ( b + ( 2 + 4 b )^2 ) x \ , \\
\psi ( x ) = 27 a \frac{ x^2 }{ ( 1- 4 x )^3 } =
27 a \frac{ - ( 2 + 4 b ) x + b }
{ -( 108 + 256b + 64 ( 2 + 4 b )^2 ) x + 1 + 48 b +
64 ( 2 + 4 b ) b  }  .
\end{array}
$$
Since $ \psi ( x_1 ) = \psi ( x_2 ) = 1$ and $x_1 \neq x_2$,
this implies
\[
0 = \left| \begin{array}{cc}
- 2 - 4 b  &  b  \\
- 108 - 256b - 64 ( 2 + 4 b )^2  & 1 + 48 b + 64 ( 2 + 4 b ) b
\end{array}
\right| = 2 ( 8 b - 1) ( 4 b + 1 ) \ ,
\]
hence
\[
b = \frac{1}{8} \ , \ \ \ \ x_i = \frac{ -5 \pm \sqrt{27} }{4} \ ,
\]
and
\[
\psi ( x ) = \frac{ ( 1 + 8x )^3}{ 64 x ( 1 - x )^3} \ .
\]
$\Box$ \par \vspace{0.2in}
Now we are in a position to prove Theorem 1.
\par \vspace{0.2in} \noindent
{\it Proof of Theorem 1. }
By Lemma 2, the map $\Psi$ in (\req(compRRJ)) and the
corresponding value of
$\alpha - \beta \ ( = 2 \alpha - 1 )$ in (\req(z2pi))
are expressed  by
\[
1728 J = 1728 \Psi ( z ) =
\left\{ \begin{array}{lll}
C \frac{ ( 1 + 8 \lambda z )^3}
{ z ( 1 - \lambda z )^3} \ , & C = \frac{1728}{ 64 \lambda } \
, & \alpha - \beta = \frac{ 1}{ 3} ; \\ [2mm]
C \frac{ ( 1 + 3\lambda z )^3}
{ z ( 1 - \lambda z )^2} \ , &
C = \frac{1728}{ 27 \lambda } \ , & \alpha - \beta =
\frac{ 1}{ 2};
\\ [2mm]
C \frac{ 1}{ z ( 1 - \lambda z )} \ , &
C = \frac{1728}{ 4 \lambda } \ , & \alpha - \beta =
\frac{ 2}{ 3} .
\end{array} \right.
\]
Since both $1728J$ and $z$ have the integral
${\bf q}$-expansions by Lemma 1, $C$ is equal to 1. Hence
\be
1728 J = 1728 \Psi ( z ) =
\left\{ \begin{array}{lll}
 \frac{ ( 1 + 216 z )^3}
{ z ( 1 - 27 z )^3} \ , & \lambda = 27 \ ,
& \alpha - \beta = \frac{ 1}{ 3} ; \\ [2mm]
 \frac{ ( 1 + 192 z )^3}
{ z ( 1 - 64 z )^2} \ , & \lambda = 64 \ , & \alpha - \beta =
\frac{ 1}{ 2} ; \\ [2mm]
 \frac{ 1}{ z ( 1 - 432 z )} \ ,  & \lambda = 432 \ ,
& \alpha - \beta =
\frac{ 2}{ 3},
\end{array} \right.
\ele(Jz)
and the corresponding operators in (\req(eqz)) are given by
$$
\begin{array}{lll}
 \lambda = 27 , & ( \alpha , \beta ) =
( \frac{2}{3}, \frac{1}{3} ) , &
\Theta^2 - 3 ( 3 \Theta + 2) (  3 \Theta + 1 ) \ ; \\ [2mm]
\lambda = 64 , & ( \alpha , \beta ) =
( \frac{3}{4}, \frac{1}{4} ) , &
\Theta^2 - 4 ( 4 \Theta + 3) (  4 \Theta + 1 ) \ ; \\ [2mm]
 \lambda = 432 , & ( \alpha , \beta ) =
( \frac{5}{6}, \frac{1}{6} ) , &
\Theta^2 - 12 ( 6 \Theta + 5) ( 6 \Theta + 1 ) \ .
\end{array}
$$
Therefore the above differential operators are the
only possibilities satisfying the conditions of Theorem 1. Now
we are going to show that the function ${\bf t} ( z )$
associated to the differential operators in Table (I) does
arise from $J$-function with the corresponding expression
of $1728J ( z )$ given there, which implies
the integral ${\bf q}$-series for $z ( {\bf q } )$ by Lemma 1.
Associated to each family of the weighted hypersurfaces
in Table (I), there corresponds an algebraic surface, denoted
again by:
$$
\begin{array}{lll}
P_8 : & x_1^3 + x_2^3 + x_3^3 - s x_1x_2x_3 = 0 ,  &
( [x_1, x_2, x_3 ], [1, s] ) \in \PZ^2 \times \PZ^1 ; \\
X_8 : & x_1^4 + x_2^4 + x_3^2 - s x_1x_2x_3 = 0 ,  &
( [x_1, x_2, x_3 ], [1, s] ) \in
\PZ^2_{(1,1,2)} \times \PZ^1 ; \\
J_{10}: &  x_1^6 + x_2^3 + x_3^2 - s x_1x_2x_3 = 0 ,  &
( [x_1, x_2, x_3 ], [1, s] )
\in \PZ^2_{(1,2,3)} \times \PZ^1 . \\
\end{array}
$$
Their singularities are given by
$$
\begin{array}{lll}
\mbox{Sing} ( P_8 ) & =  & \emptyset ; \\
\mbox{Sing} ( X_8 ) & =  & \{
( [0, 0, 1 ], [1, \infty] ) \}  ; \\
\mbox{Sing} ( J_{10}) & = &
\{ ( [0, 1, 0 ], [1, \infty] ) , ( [0, 0, 1 ], [1, \infty] )
\} .
\end{array}
$$
Let
\[
S = S ( P_8 ) , \ S ( X_8 ) , \ S ( J_{10} )
\]
be the minimal resolution of the corresponding surface, which
is an elliptic surface over $\PZ^1$ via the
$s$-projection:
\[
\sigma : S \longrightarrow \PZ^1 \ ,
\]
with the singular fiber $\sigma^{-1} ( \infty )$ of type $_1\mbox{I}_b$:

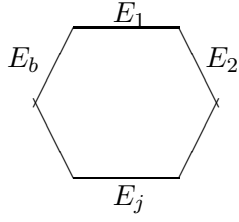
\[
\put(-20, 0){\line(1,0){40}}
\put(20, 0){\line(1,-2){15}}
\put(35, -27){\line(-1,-2){15}}
\put(-20, -57){\line(1,0){40}}
\put(-20, 0){\line(-1,-2){15}}
\put(-20, -57){\line(-1,2){15}}
\put(-5, 2){\shortstack{$E_1$}}
\put(30, -15){\shortstack{$E_2$}}
\put(-5, -67){\shortstack{$E_j$}}
\put(-45, -15){\shortstack{$E_b$}}
\]
where
\[
 b =
\left\{ \begin{array}{ll}
 3 \ , \ & S = S ( P_8 ) \ ; \\
 4 \ , \ & S = S ( X_9 ) \ ; \\
 6 \ , \ & S = S ( J_{10} ) \ .
\end{array} \right.
\]
Let $\{ \Gamma_1 , \Gamma_2 \}$ be a canonical basis of
$\mbox{H}_1 ( \sigma^{-1} ( s ) , \ZZ )$ for $  | s | \gg 0 $,
which is defined by
$$
\begin{array}{ll}
\Gamma_1 = & \mbox{the\ vanishing\ circle\ near\ a\ double\
point\ of\ } \sigma^{-1} ( \infty ) \ , \\
\Gamma_2 = & \mbox{the\ invariant\ circle\ near\ }
\sigma^{-1} ( \infty ) \ .
\end{array}
$$
The Picard-Lefschetz transformation along $s e^{i\theta}$
as $\theta$ varies from
$0$ to $- 2 \pi$ is given by
\[
( \Gamma_1, \Gamma_2 ) \mapsto
( \Gamma_1, \Gamma_2 ) \left( \begin{array}{cc}
   1 & b \\
   0 & 1   \end{array} \right) \ .
\]
Since the morphism
\[
( [ x_1 , x_2 , x_3 ] , [ 1 , s ] ) \mapsto
( [ e^{ 2 \pi i / b } x_1 , x_2 , x_3 ] , [ 1 ,
e^{ - 2 \pi i / b } s ] ) \ ,
\]
induces an order $b$ automorphism of the surface $S$, the
period map
\[
s \mapsto
(\int_{\Gamma_2} \omega_s)/(\int_{\Gamma_1} \omega_s) \ , \
\ \ \ \omega_s = \mbox{the \ holmorphic\ differential\ of\ }
\sigma^{-1}(s) \ ,
\]
is determined by the variable $z (: = s^{-b} )$. Hence we obtain
a ( multi-valued ) function ${\bf t} ( z )$ from $z$-plane
to the upper half-plane $\HZ$, which is the solution of
corresponding differential equation in Table (I),
( for its derivation , see e.g. \cite{KLRY} ).
As $z$ varies along the path $z e^{i\theta}$ from
$\theta = 0$ to $\theta = 2 \pi$, the change of
homology of the fiber in $S$ is described by
\[
( \Gamma_1, \Gamma_2 ) \mapsto
( \Gamma_1, \Gamma_2 ) \left( \begin{array}{cc}
   1 & 1 \\
   0 & 1   \end{array} \right) \ .
\]
Therefore ${\bf t} ( z )$ satisfies the condition
(\req(initial)). The function ${\bf t} ( z )$ describes the periods
of elliptic fibers of the surface $S$, hence arises
from $J$-function and we have the diagram (\req(compRRJ)).
By Lemma 2 and (\req(Jz)), we obtain the expression of
$1728 J ( z )$. This completes the proof of Theorem 1.
$\Box$ \par \vspace{0.2in} \noindent

{}From the relation of variables $J$ and $z$ in Table (I),
the map ${\bf t}$ in (\req(compRRJ)) is bijective on the region
$\mbox{Im}({\bf t}) \gg 0$ . From (\req(swt)) and the
values of $\alpha - \beta$ in (\req(Jz)), ${\bf t}$ is also bijective
on a connected region of $\Re$ over $| z | \gg 0$ for the
the cases of $P_8$ and $X_9$, but not for $J_{10}$. For the
families of $P_8$ and $X_9$,
$\overline{\Re}$ is indeed isomorphic to the upper-half
plane $\HZ$ via the map ${\bf t}$. In the following sections,
we are going to express $z ( {\bf t} ) $
in terms of theta functions without appealing to the
$J$-function.

For $J_{10}$-family, the Riemann surface $\overline{\Re}$
is not isomorphic to $\HZ$. However, from the relation of $J$ and
$z$, one can easily obtain
\[
z ( \tau ) = \frac{1 - \sqrt{ 1 - \frac{1}{J ( \tau )} }}{864}
= \frac{\sqrt{g_2( \tau )^3} - \sqrt{27} g_3(\tau)^2 }
{864 \sqrt{g_2( \tau )^3}}
\]
where $g_2, g_3$ are given by (\req(Jg23)). For an elliptic
curve in $J_{10}$-family:
$$
\begin{array}{lll}
 (J_{10}) \ \ &  X_s : \
x_1^6 + x_2^3 +x_3^2 - s x_1 x_2 x_3 = 0 , &
[x_1, x_2, x_3 ] \in {\PZ}^2_{(1,2,3)} ,
\end{array}
$$
the substitution,
\be
i y =   \frac{2 x_3 -  s x_2x_1}{x_1^3}
,  \ \ x = \frac{x_2 - \frac{s^2}{12}x_1^2}{x_1^2} \ ,
\ele(J10W)
changes $X_s$ to the Weierstrass form (\req(Weierf)) with
\[
g_2 = \frac{ s^4 }{12} \ , \ \ g_3 = \frac{ s^6 }{216} - 4 \ .
\]
By using the homogeneous coordinates of ${\PZ}^2_{(1,2,3)}$,
\[
[ x_1, x_2, x_3 ] = [ 1 , x , i y ] \ ,
\]
an elliptic curve of Weierstrass form (\req(Weierf)) is
expressed by
\[
x_3^2 + 4 x_3^3 - g_3 x_1^6  - g_2 x_1^4 x_2   = 0 \ , \
\ [ x_1, x_2, x_3 ] \in {\PZ}^2_{(1,2,3)} ,
\]
Through Weierstrass function presentation of
(\req(Weierf)) and the formulae in \cite{WW}, one obtains
the theta function expression of the above $x_i$'s :
$$
\begin{array}{lll}
x_1 & = & 2 \vartheta_1 ( z, \tau ) \ , \\
x_2 & = & \frac{1}{3}
(\vartheta_3^4 ( 0, \tau ) + \vartheta_3^4 ( 0, \tau ) )
\vartheta_1^2 ( z, \tau ) + \vartheta_3^2 ( 0, \tau )
\vartheta_4^2 ( 0, \tau ) \vartheta_2^2 ( z, \tau ) \ , \\
x_3 & = & -2 i \vartheta_2^2 ( 0, \tau )\vartheta_3^2 ( 0, \tau )
\vartheta_4^2 ( 0, \tau )\vartheta_2(z, \tau)
\vartheta_3(z, \tau)\vartheta_4(z, \tau) \ ,
\end{array}
$$
with $J = \frac{g_2^3}{g_2^3 - 27 g_3^2}$ given by
(\req(Jg23)).
In a similar way, one can obtain the theta function
representation for $J_{10}$-family
via (\req(J10W)).

\section{ Theta Function Parametrization for $P_8$-family }
In this section we describe the elliptic theta function
representation of elliptic curves in $P_8$-family,
\bea(lll)
( P_8 ) \ \ \  & \ X_s :  f_s ( x ) =
x_1^3 + x_2^3 + x_3^3 - s x_1 x_2 x_3 = 0 , &
[x_1, x_2, x_3 ] \in {\PZ}^2 .
\elea(eqXs)
The moduli parameter $z \ ( : = s^{-3} )$ will be expressed
by theta constants involved in the representation.
First we note that the fundamental locus of the  pencil
(\req(eqXs)) consists of 9 elements:
\be
[x_1, x_2, x_3] = [ 0, -1 , \omega^k ] ,
[ -\omega, 0 , -1] , [ -1 , \omega^k, 0 ] \ , \ \mbox{for} \ 0 \leq k \leq 2 \
, \
\omega = e^{ \frac{2 \pi i}{3} } \ ,
\ele(9fund)
each of which gives rise to a section of the ellpitic surface:
\be
\sigma : S ( P_8 ) \longrightarrow \PZ^1 \ .
\ele(SP8)
Let $G$ be the group  of linear transformations
preserving the polynomal $f_s ( x )$ for a generic $s$. It is
easy to see that the finite group $G$
is generated by the following elements:
\bea(ll)
C ( x_1, x_2 , x_3 ) = ( \omega x_1 , \omega x_2 , \omega x_3 ) \ , \ &
R ( x_1, x_2 , x_3 ) = ( x_1 , \omega x_2 , \omega^2 x_3 ) \ \\
T ( x_1, x_2 , x_3 ) = ( x_2 , x_3 , x_1 ) \ , \ &
I ( x_1, x_2 , x_3 ) = ( x_1 , x_3 , x_2 ) \ .
\elea(WRTI)
The subgroup generated by $C, R$ is described by
\[
< C , R > = \{ \mbox{dia.} ( \alpha_1, \alpha_2, \alpha_3 ) \
| \ \alpha_1^3 = \alpha_2^3 = \alpha_3^3 = \alpha_1 \alpha_2 \alpha_3 = 1 \} \
{}.
\]
By the relation
\[
R \cdot T = C \ ( T \cdot R ) \ ,
\]
one can easily see that $G$ is isomorphic to the extended
degree 3 Heisenberg group $\widetilde{\GZ}_3$:
\be
G \cap SL_3 ( \CZ ) = < C , R , T > \simeq \GZ_3 \ , \ \
G  = < C , R , T > \cdot < I > \simeq \widetilde{\GZ}_3 \ .
\ele(GG3)
The action of $G$ on the homogeneous coordinates
$x_k$'s is now equivalent to the canonical representation of
$\widetilde{\GZ}_3$ by identifying $x_k$ with
$e_k$ of (\req(canrep)).
As a projective transformation group,
$G$ acts on ${\PZ}^2$ which leaves each $X_s$ invariant.
Let $r_s , t_s , \iota_s$ be the automorphisms of $X_s$
induced by $R, T, I$ respectively. Then
$t_s$ and $r_s$ generates the group of
order 3 translations of $X_s$, and $\iota_s$ is an
involution of $X_s$:
\bea(clcl)
< r_s , t_s  > & \simeq & \ZZ_3 \oplus \ZZ_3  &
\simeq
\ \GZ_3 / \mbox{center}( \GZ_3 )  \ , \ \\
< r_s , t_s , \iota_s  > & \simeq &
(\ZZ_3 \oplus \ZZ_3 ) \cdot \ZZ_2  & \simeq
\ \widetilde{\GZ}_3 / \mbox{center}( \widetilde{\GZ}_3 )  \ .
\elea(G3Xs)
The fundamental locus (\req(9fund)) is
invariant under $\iota_s$ with only one fixed element
$[ 0 , -1, 1]$, which
induces a section of the elliptic surface (\req(SP8)), denoted
by
\[
\rho : \PZ^1 \longrightarrow S ( P_8 ) \ , \ s \mapsto
\rho ( s ) \ .
\]
All the 9 sections induced by (\req(9fund)) are the
translations of the above $\rho$ by elements in
$< t_s , r_s >$. Denote
\[
{\cal O}_{X_s} ( 1 ) = \mbox{the\ restriction\ of\ hyperplane\
bundle\ on\ } X_s \ .
\]
By (\req(GG3)) and (\req(G3Xs)), we have a
$\widetilde{\GZ}_3$-linearization on the line bundle
${\cal O}_{X_s} ( 1 )$ via the linear representation of $G$ ,
\bea(clc)
\widetilde{\GZ}_3 \times  {\cal O}_{X_s} ( 1 ) & \longrightarrow &
{\cal O}_{X_s} ( 1 )    \\
\downarrow & & \downarrow  \\
\widetilde{\GZ}_3 \times  X_s & \longrightarrow & X_s \ .
\elea(G3O)
We are going to construct this
$\widetilde{\GZ}_3$-linearization from the universal covering
space of $X_s$.
It is known that he quotient of
${\PZ}^2 $ by $ <R> $ is a 2-dimensional toric variety :
\[
{\PZ}^2 / <R> \ \supset \
{ \CZ }^{*3}/ ( { \CZ }^{*}\cdot <R> ) \ ,
\]
and the natural projection defines an equivariant morphism
of toric varieties,
\[
p : {\PZ}^2  \longrightarrow {\PZ}^2 /<R> \ .
\]
There are 3 toric divisors in ${\PZ}^2/<R> $, given by
$\xi_i = 0$ for $1 \leq i \leq 3$, here $\xi_i$'s are
sections on ${\PZ}^2 /<R>$ with $p^*( \xi_i ) = x_i $.
Note that $\xi_i$
and $\xi_j$ are not linearly equivalent for $ i \neq j$, however $\xi_i^3$'s
and $\xi_1 \xi_2 \xi_3$ are sections of the same line bundle.
Denote
\be
\Xi_s :  \xi_1^3 + \xi_2^3 + \xi_3^3 - s \xi_1 \xi_2 \xi_3 = 0
\ \ \ \ \mbox{in \ } {\PZ}^2 /<R> \ .
\ele(eqxi)
The restriction of the projection $p$
defines a 3-fold cover of elliptic curves:
\[
p_s : X_s \longrightarrow \Xi_s = X_s / < r_s > \ ,
\]
and the automorphisms $\iota_s , t_s$ of $X_s$ induce
the involution $\iota_{s, 0}$ and an order 3
translation $t_{s, 0}$ of $\Xi_s$ respectively. The restriction
of $p_s$ defines an one-one correspondence between the fixed
points $\iota_s$ and $\iota_{s, 0}$,
\[
p_s : X_s^{\iota_s} \ \simeq \ \Xi_s^{ \iota_{s,0} } \ ,
\]
and denote the element in $\Xi_s^{ \iota_{s,0} }$
corresponding to $\rho ( s )$ by
\[
\rho ( s )_0  : = p_s ( \rho ( s )) = \mbox{zero}
( \xi_1 ) .
\]
One may regard $\Xi_s$ as an 1-dimensional torus $\EZ_{\tau,1}$
for some $ \tau \in \HZ$ with
\bea(cll)
\iota_{s,0} : \Xi_s \longrightarrow \Xi_s ,
 & \Longleftrightarrow &
\iota : \EZ_{\tau,1} \longrightarrow \EZ_{\tau,1} , \ \
 [{\rm z}] \mapsto [-{\rm z}] \\
t_{s,0} : \Xi_s \longrightarrow \Xi_s
& \Longleftrightarrow &
t : \EZ_{\tau,1} \longrightarrow \EZ_{\tau,1} , \
\ [{\rm z}] \mapsto [{\rm z + c}] \ , \ \ \mbox{for \ }
[{\rm c}] \in
\EZ_{\tau,1} ( 3 ) , \\
\rho ( s )_0  \in \Xi_s & \Longleftrightarrow
& e \in \EZ_{\tau,1} ( 2 ) \ .
\elea(XiE)
The above data indeed determine the algebraic form
(\req(eqxi)) by the following lemma.
\par \vspace{0.2in} \noindent
{ \bf Lemma 3 . } Let $\EZ$ be an elliptic curve with
an involution $\iota$ and an order 3 translation $t$.
Let $e$ be an element of $\EZ$ fixed by $\iota$. Then:
\par \noindent
(i) The following divisors are linearly equivalent:
\[
e + t ( e ) + t^2 ( e ) \sim \ 3e \sim \ 3 t ( e ) \sim \
3 t^2 ( e ) .
\]
(ii) There exist non-trivial sections $f_i$
in $\Gamma ( \EZ , {\cal O} ( t^{i-1} ( e ) )$ ,
 $1 \leq i \leq 3$ , such that
\[
 f_1^3 +  f_2^3  + f_3^3  = s f_1 f_2 f_3 \ \ \
\ \in \Gamma ( \EZ , {\cal O} ( 3 e  ) ) \ ,
\]
for some $s \in \CZ - \{ 0 \}$.
\par \vspace{0.2in} \noindent
{\it Proof.} Since
\[
\iota ( t (e) ) = (\iota t \iota ) ( e ) = t^{-1} ( e ) =
t^2 ( e ) ,
\]
we have
\[
2 e \sim t ( e ) + t^2 ( e ) \ ,
\]
hence
\[
3 e \sim e + t ( e ) + t^2 ( e ) \ .
\]
Applying $t$ and $t^2$ to the above relation, we obtain (i).
For $1 \leq i \leq 3$, let $f_i$ be a non-trivial element in
$\Gamma ( \EZ , {\cal O} ( t^{i-1} ( e ) )$.
Since $\Gamma ( \EZ , {\cal O} ( 3 e  ) )$ is a 3-dimensional
vector space with $\{ f_i^3 \}_{i=1}^3 $ as a basis, we have
the relation:
\[
f_1 f_2 f_3 =  \beta_1 f_1^3 + \beta_2 f_2^3 + \beta_3 f_3^3  \ ,
\]
for $\beta_i \in \CZ$. As $e , t ( e )$ and $t^2 ( e )$ are
distinct elements in $\EZ$, one concludes $\beta_i \neq 0$ for all $i$.
Replacing $f_i$ by $\beta_i^{\frac{1}{3}} f_i$, we obtain (ii).
$\Box$ \par \vspace{0.2in} \noindent
One may describe the above $f_i$'s by theta
functions with characteristics on an
elliptic curve $\EZ_{\tau, 1}$, and then the equation in (ii)
simply means the relation among those theta functions. A
such relation is given as follows.
\par \vspace{0.2in} \noindent
{ \bf Lemma 4. } For $\tau \in \HZ$, let $p_1, p_2, p_3$ be
the elements in  $\EZ_{\tau, 1}$ defined by
\[
p_1 = [ \frac{ \tau }{2} + \frac{ 1}{2} ] \ , \
p_2 = [ \frac{ \tau }{6} + \frac{ 1}{2}  ] \ , \
p_3 = [ \frac{ 5 \tau }{6} + \frac{ 1}{2}  ] \ .
\]
Let
\[
\xi_1 = \vartheta [ \begin{array}{c} 0 \\ 0
\end{array} ] ( {\rm z} , \tau ) \ , \
\xi_2 = \vartheta [ \begin{array}{c} \frac{1}{3} \\ 0
\end{array} ] ( {\rm z} , \tau ) \ , \
\xi_3 = \vartheta [ \begin{array}{c} \frac{2}{3} \\ 0
\end{array} ] ( {\rm z} , \tau )   .
\]
Then
\[
\xi_1^3 \ ,  \ \xi_2^3 \ ,  \ \xi_3^3 \ ,  \
\xi_1\xi_2\xi_3 \ \in \Gamma (
\EZ_{\tau, 1} , {\cal O}( \sum_{i =1}^3 p_i ) )
\]
and the following relation holds:
\be
\xi_1^3  + \xi_2^3 + \xi_3^3 = s  \xi_1\xi_2\xi_3 \ ,
\ele(xi3prod)
with $s$ given by
\[
s^{-1} =
\frac{ q^{5/9}\vartheta ( 0, \tau ) \vartheta ( \frac{ \tau}{3} , \tau )
\vartheta ( \frac{ 2 \tau}{3} , \tau )}
{\vartheta ( 0, \tau )^3 + q^{1/3}\vartheta ( \frac{ \tau}{3} , \tau )^3
+ q^{4/3} \vartheta ( \frac{ 2 \tau}{3} , \tau )^3 } \ .
\]
\par \vspace{0.2in} \noindent
{\it Proof.}
By (\req(prthetacha)), the
zeros of $  \xi_1\xi_2\xi_3 $ are $ p_1 + p_2 + p_3 $,
and the functions $
\xi_1^3, \xi_2^3 , \xi_3^3 , \xi_1\xi_2\xi_3$ have the same
quasi-periodicity condition,
hence define four sections in $\Gamma (\EZ_{\tau, 1} ,
{\cal O}( \sum_{ i=1}^3 p_i ) )$. Now the linear
dependence of $\xi_1^3  + \xi_2^3 + \xi_3^3 $ and
$ \xi_1\xi_2\xi_3 $ is equivalent to
\[
\mbox{zero  } ( \xi_1^3  + \xi_2^3 + \xi_3^3 ) =
p_1 + p_2 + p_3 \ ,
\]
which will follow from
\[
( \xi_2^3 + \xi_3^3 ) ( p_1 ) =
( \xi_1^3 + \xi_3^3 ) ( p_2 ) =
( \xi_1^3 + \xi_2^3 ) ( p_3 ) = 0 .
\]
By (\req(prthetacha)) (\req(-zzero)), we have
$$
\begin{array}{llll}
\vartheta
[ \begin{array}{c} \frac{2}{3} \\ 0
\end{array} ] ( \frac{ \tau + 1}{2} , \tau ) &=&
- e^{ 2 \pi i / 3} \vartheta
[ \begin{array}{c} \frac{1}{3} \\ 0
\end{array} ] ( \frac{ \tau + 1}{2} , \tau ) \ &; \\
\vartheta
[ \begin{array}{c} \frac{2}{3} \\ 0
\end{array} ] ( \frac{ \tau + 1}{2} - \frac{\tau}{3}, \tau )
&=& e^{2 \pi i \frac{\tau + 1}{3}} \vartheta (
\frac{ \tau + 1}{2} + \frac{\tau}{3} )& =
e^{2 \pi i \frac{\tau + 1}{3}} \vartheta ( -
\frac{ \tau + 1}{2} - \frac{\tau}{3} ) \\
&=&  - e^{2 \pi i / 3 } \vartheta (
\frac{ \tau + 1}{2} - \frac{\tau}{3} )& ; \\
\vartheta
[ \begin{array}{c} \frac{1}{3} \\ 0
\end{array} ] ( \frac{ \tau + 1}{2} - \frac{2 \tau}{3}, \tau )
&=& e^{ \pi i /3 } \vartheta (
\frac{ \tau + 1}{2} - \frac{\tau}{3} )& =
e^{ \pi i /3 } \vartheta ( -
\frac{ \tau + 1}{2} + \frac{\tau}{3} ) \\
&=&  e^{ \pi i / 3 } \vartheta (
\frac{ \tau + 1}{2} - \frac{2\tau}{3} )& . \\
\end{array}
$$
Therefore we obtain the relation (\req(xi3prod)), whose value
at ${\rm z} = 0$ gives the expression of $s$ .
$\Box$ \par \vspace{0.2in} \noindent
{ \bf Remark.} By a similar argument, one can also have the
relation (\req(xi3prod)) by setting
\bea(llll)
\xi_1 = \vartheta [ \begin{array}{c} 0 \\ 0
\end{array} ] ( {\rm z} , \tau ) \ , &
\xi_2 = \vartheta [ \begin{array}{c} 0 \\ \frac{1}{3}
\end{array} ] ( {\rm z} , \tau ) \ , &
\xi_3 = \vartheta [ \begin{array}{c} 0 \\ \frac{2}{3}
\end{array} ] ( {\rm z} , \tau ) ;
\\ [4mm]
\xi_1 = \vartheta [ \begin{array}{c} 0 \\ 0
\end{array} ] ( {\rm z} , \tau ) \ , &
\xi_2 = e^{8 \pi i/ 9} \vartheta [ \begin{array}{c} \frac{1}{3} \\ \frac{1}{3}
\end{array} ] ( {\rm z} , \tau ) \ , &
\xi_3 = e^{8 \pi i/ 9} \vartheta [ \begin{array}{c} \frac{2}{3} \\ \frac{2}{3}
\end{array} ] ( {\rm z} , \tau ) ; \\[4mm]
\xi_1 = \vartheta [ \begin{array}{c} 0 \\ 0
\end{array} ] ( {\rm z} , \tau ) \ , &
\xi_2 = e^{- 2 \pi i/ 9} \vartheta [ \begin{array}{c} \frac{1}{3} \\
\frac{2}{3}
\end{array} ] ( {\rm z} , \tau ) \ , &
\xi_3 = e^{- 2 \pi i/ 9} \vartheta [ \begin{array}{c} \frac{2}{3} \\
\frac{1}{3}
\end{array} ] ( {\rm z} , \tau ) ;
\elea(1/3)
with the corresponding $s$ given by
$$
\begin{array}{ll}
s^{-1} = &
\frac{ \vartheta ( 0, \tau ) \vartheta ( \frac{1}{3} , \tau )
\vartheta ( \frac{ 2 }{3} , \tau )}
{\vartheta ( 0, \tau )^3 + \vartheta ( \frac{1}{3} , \tau )^3
+  \vartheta ( \frac{ 2 }{3} , \tau )^3 } \ ; \\[2mm]
s^{-1} = &
\frac{  q^{5/9} e^{ 8 \pi i / 9 }\vartheta ( 0, \tau )
\vartheta ( \frac{ \tau + 1 }{3} , \tau )
\vartheta ( \frac{ 2 \tau + 2 }{3} , \tau )}
{\vartheta ( 0, \tau )^3 + q^{1/3} e^{- 2 \pi i/ 3 }
\vartheta ( \frac{ \tau + 1 }{3} , \tau )^3
+ q^{4/3} e^{ - 2 \pi i/ 3 } \vartheta ( \frac{ 2 \tau + 2 }{3} ,
\tau )^3 } \ ;
\\ [2mm]
s^{-1} = &
\frac{  q^{5/9} e^{ 4 \pi i / 9 }\vartheta ( 0, \tau )
\vartheta ( \frac{ \tau + 2}{3} , \tau )
\vartheta ( \frac{ 2 \tau + 1 }{3} , \tau )}
{\vartheta ( 0, \tau )^3 + q^{1/3} e^{ 2 \pi i/ 3 }
\vartheta ( \frac{ \tau + 2}{3} , \tau )^3
+ q^{4/3} e^{ 2 \pi i/ 3 }
\vartheta ( \frac{ 2 \tau + 1}{3} , \tau )^3 } \ .
\end{array}
$$

With the elliptic curve $\Xi_s$ identified with
$\EZ_{\tau, 1}$ as in (\req(XiE)), one can write
$X_s = \CZ / L $ for an index 3 sublattice $L$
of $\ZZ \tau + \ZZ$.  With the complex number c in
(\req(XiE)), one may assume that
\[
\CZ \longrightarrow \CZ \ , \ \ \ {\rm z} \mapsto  {\rm z + c} \ ,
\]
induces the order 3 automorphism $t_s$ of $X_s$. Then one can
easily conclude
\be
( X_s , < t_s > , < r_s > ) = \left\{ \begin{array}{ll}
( \EZ_{ \tau , 3} \ , < [{\rm z}] \mapsto [{\rm z} + \frac{\tau}{3}] > ,
< [{\rm z}] \mapsto [{\rm z} + 1] > ) & \mbox{if\ } {\rm c} =
\pm \frac{ \tau}{3} , \\
( \EZ_{ \tau + 1 , 3} \ , < [{\rm z}] \mapsto [{\rm z} + \frac{\tau + 1}{3}] >
,
< [{\rm z}] \mapsto [{\rm z} + 1] > ) & \mbox{if\ } {\rm c} =
\pm \frac{\tau + 1}{3} , \\
( \EZ_{ \tau + 2 , 3} \ , < [{\rm z}] \mapsto [{\rm z} + \frac{\tau + 2}{3}] >
,
< [{\rm z}] \mapsto [{\rm z} + 1] > ) & \mbox{if\ } {\rm c} = \pm \frac{\tau +
2}{3} , \\
( \EZ_{ 3\tau , 1} \ , < [{\rm z}] \mapsto [{\rm z} + \frac{1}{3}] > ,
< [{\rm z}] \mapsto [{\rm z} + \tau] > ) & \mbox{if\ } {\rm c} = \pm
\frac{1}{3}  .
\end{array}
\right.
\ele(XsE)
Consider $\xi_i$ and $x_i$ as functions on the universal
covering space $\CZ$ of $\Xi_s$, and regard the fundamental
group of $\Xi_s$ as a subgroup of the Heisenberg group $\GZ$,
which acts on entire functions of $\CZ$ as in Sect. 2. By
(\req(XsE)) and (\req(WRTI)), $\xi_1$ corresponds to the
$\Lambda ( 1, 1)$-entire function on $\CZ$, hence we have
\[
\xi_1 = \vartheta [ \begin{array}{c} 0 \\ 0
\end{array} ] ( {\rm z} , \tau ) \ .
\]
By Lemma 3, renumbering  $\xi_2$ and $\xi_3$ if necessary,
one may represent $\xi_i$'s by theta functions
either in Lemma 4 , or those in (\req(1/3)).
By (\req(WRTI)) (\req(XiE)), one requires
\[
\xi_2 ( - {\rm z} ) \ = \ \xi_3 ( {\rm z} ) \ ,
\]
hence by (\req(-zzero)), we have
\[
\xi_2 ( {\rm z} ) = \vartheta [ \begin{array}{c} \frac{1}{3} \\ 0
\end{array} ] ( {\rm z} , \tau ) \ \ \ \mbox{or\ }
\ \ \vartheta [ \begin{array}{c} 0 \\ \frac{1}{3}
\end{array} ] ( {\rm z} , \tau ) \ .
\]
By (\req(WRTI)) (\req(XsE)), one has the relation:
\[
\xi_2 ( {\rm z} + 1 ) \ = \ \omega \xi_2 ( {\rm z}) \ ,
\]
therefore
\[
\xi_2 ( {\rm z} ) = \vartheta [ \begin{array}{c} \frac{1}{3} \\ 0
\end{array} ] ( {\rm z} , \tau ) \ .
\]
Then Lemma 4 implies the following result:
\par \vspace{0.2in} \noindent
{ \bf Theorem 2. } With $x_i ,
\xi_i \ ( 1 \leq i \leq 3)$ the coordinates of
elliptic curves $X_s , \Xi_s$ in $P_8$-family, $p_s$ the morphism between
them, and $ r_s , t_s , \iota_s $ the automorphisms of $X_s$
as before.
For $\tau \in \HZ$, define $s ( \tau )$ by
\[
s ( \tau )^{-1} =
\frac{ q^{5/9}\vartheta ( 0, \tau ) \vartheta ( \frac{ \tau}{3} , \tau )
\vartheta ( \frac{ 2 \tau}{3} , \tau )}
{\vartheta ( 0, \tau )^3 + q^{1/3}\vartheta ( \frac{ \tau}{3} , \tau )^3
+ q^{4/3} \vartheta ( \frac{ 2 \tau}{3} , \tau )^3 } \ .
\]
Then the above data for $X_s , \Xi_s$ have the following
realization in complex tori:
$$
\begin{array}{lll}
X_s = \EZ_{ \tau , 3} \ , & \Xi_s = \EZ_{ \tau , 1} \ , &
 p_s : X_s \longrightarrow \Xi_s \ , \ [{\rm z}] \mapsto
[{\rm z}] \ ;
\\ [2mm]
\xi_1 = \vartheta [ \begin{array}{c} 0 \\ 0
\end{array} ] ( {\rm z} , \tau ) \ , &
\xi_2 = \vartheta [ \begin{array}{c} \frac{1}{3} \\ 0
\end{array} ] ( {\rm z} , \tau ) \ ,&
\xi_3 = \vartheta [ \begin{array}{c} \frac{2}{3} \\ 0
\end{array} ] ( {\rm z} , \tau ) \ ; \\ [2mm]
 t_s : \EZ_{ \tau , 3} \longrightarrow \EZ_{ \tau , 3} \ ,&
[{\rm z}] \mapsto [{\rm z} + \frac{\tau}{3}] \ ; \\
 r_s : \EZ_{ \tau , 3} \longrightarrow \EZ_{ \tau , 3} \ ,&
 [{\rm z}] \mapsto [{\rm z} + 1]  \ ; \\
\iota_s : \EZ_{ \tau , 3} \longrightarrow \EZ_{ \tau , 3} \ ,&
[{\rm z}] \mapsto [-{\rm z}] ,
\end{array}
$$
and the projective representation of $< r_s , t_s , \iota_s> $
on $x_i$'s is given by the canonical representation of
$\widetilde{\GZ}_3$ on $\mbox{Th}_3 ( \tau )$.
$\Box$ \par \vspace{0.2in} \noindent
By the above expression of $s ( \tau )^{-1}$ , we now
derive the formula of the variable $z \ (:= s^{-3})$ as an
function of ${\bf q}$ in the following theorem.
\par \vspace{0.2in} \noindent
{\bf Theorem 3 . } The function $z ( {\bf q} )$ for
$P_8$-family  of
Table (I) is given by
\[
z ( {\bf q} ) = \frac{ {\bf q}^{5/2} \vartheta ( 0, 3 {\bf t} )^3
\vartheta (  {\bf t} , 3 {\bf t} )^3
\vartheta ( 2 {\bf t} , 3 {\bf t} )^3 }
{ ( \vartheta ( 0, 3 {\bf t} )^3 +
{\bf q}^{1/2} \vartheta (  {\bf t} , 3 {\bf t} )^3
+ {\bf q}^2 \vartheta ( 2 {\bf t} , 3 {\bf t} )^3 )^3 }  \ ,
\ \ {\bf q} = e^{ 2 \pi i {\bf t} } \ ,
\]
and it has the integral ${\bf q}$-expansion with
\[
\lim_{{\bf q} \rightarrow 0 }
\frac{z ({\bf q} )}{ {\bf q}} = 1 \ .
\]
\par \vspace{0.2in} \noindent
{\it Proof. } By Theorem 2, for $s = s ( \tau )$,
we have
\[
X_s = \EZ_{ \tau , 3} \simeq \EZ_{ {\bf t} , 1} \ , \
\  {\bf t} = \frac{\tau}{3} \ ,
\]
then one obtains the above expresson of $z ({\bf q})$.
By the infinite $q$-product representation of the theta function,
the ratio $\frac{z ({\bf q} )}{ {\bf q}}$ tends to 1 as
${\bf q} \rightarrow 0$, and
$z ( {\bf q} ) $ is an integral
power series of $\sqrt{\bf q} $. The variable
${\bf t}$ is obtained as the ratio of two periods of the holomorphic
differential of $X_s$. As $X_s$ is isomorphic to
$X_{\omega s}$, ${\bf t}$ can be considered as a multi-valued
function of $z$. Since the periods satisfy the Fuchsian
equation (\req(eqz)) for $\lambda = 27,  \alpha = \frac{2}{3} ,
\beta = \frac{1}{3}$ , ${\bf t}$ is a solution
of the corresponding Schwarzian equation (\req(scheqn)). It can
be shown that
${\bf t} ( z )$ satisfies the condition (\req(initial)),
hence ${\bf t}$ is the variable in Sect. 3. Since $z$ is a
function of ${\bf q}$, this implies  $z ( {\bf q} ) $ is a
power series of ${\bf q}$ with integral coefficients expressed
in (\req(qexp)).
$\Box$ \par \vspace{0.2in} \noindent
{\bf Remark.}

(i) In this expression of $z ({\bf q})$,
both the numerator and denominator are integral
power series of $\sqrt{\bf q} $, but not in ${\bf q}$.
However their ratio gives an integral ${\bf q}$-expansion
for $z$.

(ii) Since the surface $S ( P_8 )$ of (\req(SP8)) is the universal
family of $( \EZ , \EZ(3) )$ for 1-torus $\EZ$  with
3-torsion $\EZ(3)$,  the correspondence
\[
\tau \mapsto s^{-1} ( \frac{ \tau }{3} ) \ ,
\]
defines an isomorphism between $\Gamma (3)\setminus\HZ $ and
$\PZ^1$.

\section{ Elliptic Curves in Ising Model }
Now we start to investigate the relation between the
constrained polynomials of $X_9$-family
and the Boltzmann weights in Ising model. Let us first recall the
Jacobi elliptic function parametrization in Ising model.
This theory has been extensively discussed in many literatures
, e.g. \cite{AP} \cite{B} \cite{TF79}. Here we adopt the
formulation in chiral Potts $N$-state model models
\cite{Bax} \cite{R92} , even though prime
interests of which were on hyperelliptic curves for $N \ge 3$, however the
parametrization works also for $N=2$, i.e. the case of
Ising model.

Let $W$ be an 1-dimensional torus ( = $\CZ / \mbox{lattice}$ )
, and consider the morphisms of $W$,
$$
\begin{array}{lll}
\theta : W \longrightarrow W , & [{\rm z}] \mapsto [-{\rm z}] &; \\
m : W \longrightarrow W , & [{\rm z}] \mapsto [ {\rm z} +
{\rm z}_0 ] , & [{\rm z}_0] \in E(2) \ ; \\
\sigma : W \longrightarrow W , &  \sigma = m \cdot \theta &.
\end{array}
$$
One can present $W$ as a plane curve through the following
commutative diagram:
\bea(lllll)
& W & \stackrel{\Psi}{\longrightarrow} & \PZ^1 & = W/ <\theta> \\
& \downarrow \Pi & & \downarrow \pi & \\
W/<\sigma>=& \PZ^1 &\stackrel{\psi}{\longrightarrow} &
\PZ^1 & = W/ <\theta, \sigma > \ ,
\elea(CPM2)
where $\Psi , \psi, \Pi , \pi$ are natural projections.
For some suitable coordinates of $\PZ^1 $,
$\psi , \pi$ have the expressions:
\[
\psi ( t ) = t^2 \ , \ \ \pi ( \lambda ) =
\frac{(1-k^\prime \lambda ) ( 1 - k^\prime \lambda^{-1} )
}{k^2} \ \ , \ \ \ \ t , \lambda \in \PZ^1 = \CZ \cup \{ \infty \} \ ,
\]
where $k^\prime , k $ are elements in $\CZ - \{ 0 , \pm 1 \}$ satisfying
the relation:
\[
k^2 + k^{\prime 2} = 1 \ .
\]
Then $W$ is isomorphic to the algebraic curve:
\be
W_{k^\prime} : \ \ t^ 2 = \frac{(1-k^\prime \lambda ) ( 1 - k^\prime
\lambda^{-1} )
}{k^2} \ , \ \ \ ( t , \lambda ) \in \CZ^2 \ ,
\ele(N=2)
with
\[
\Psi ( t , \lambda ) = \lambda \ , \ \
\Pi ( t , \lambda ) = t \ , \ \
\theta ( t , \lambda ) = ( - t , \lambda ) \ , \ \
\sigma ( t , \lambda ) =(t , \lambda^{-1} ) \ .
\]
The branched data of $\Psi$ and $\Pi$ are given by
\bea(ll)
\mbox{Branched \ points \ of\ } \Psi : &
p = (\infty ,  0 ), \  p^\prime= ( \infty , \infty ) , \
q= ( 0 , k^\prime ) , \ q^\prime = (0 , k^{\prime -1} ) \\
\mbox{Branched \ points \ of \ } \Pi :&
b_{\pm} = ( \pm \sqrt{\frac{1-k'}{1+k'}} , 1 ) , \ \
b_{\pm}^\prime = (\pm  \sqrt{\frac{1+k'}{1-k'}} , -1) \ .
\elea(branch)
By the transformations,
\[
w = \frac{k'}{k^2} ( \lambda - \frac{1}{\lambda} ) \ , \ \
\lambda = \frac{1}{2k'} \{ k^2( w - t^2) + k'^2 + 1 \} \ ,
\]
$W_{k^\prime}$ is birationally equivalent to the plane curve :
\be
w^2 = ( t^2 - \frac{1- k'}{1+k'} )( t^2 - \frac{1+ k'}{1-k'} )
\ , \ \ \ ( t , w ) \in \CZ^2 \ .
\ele(eqnw)
With $\PZ^2_{(1,1,2)}$ as a compactification of $\CZ^2$  via
the identity:
\be
[ 1 , t , i w ] = [ y_1 , y_2 , y_3 ] \in
\PZ^2_{(1,1,2)} \ ,
\ele(twyi)
the equation (\req(eqnw)) can be rewritten as
\be
W_{k^\prime} \ \simeq \ Y_{\epsilon} :
y_1^4 + y_2^4 + y_3^2 - \epsilon y_1^2 y_2^2 = 0 \ , \ \
[y_1, y_2, y_3 ] \in \PZ^2_{(1,1,2)} \ , \ \ \ ( \
\epsilon = 2\frac{1+ k'^2}{1-k'^2} \ ) \ .
\ele(Ising)
Now the branched points of $\Psi$ are given by zeros of $y_1$
or $y_2$:
$$
\begin{array}{lllll}
p : & (t , \lambda ) = ( \infty , 0 ) &
\longleftrightarrow & [ y_1 , y_2 , y_3 ] =
[ 0 , 1 , i ] ; \\
p' : & (t , \lambda ) = ( \infty , \infty ) &
\longleftrightarrow & [ y_1 , y_2 , y_3 ] =
[ 0 , 1 , - i ] ; \\
q : & (t , \lambda ) = ( 0 , k' ) &
\longleftrightarrow & [ y_1 , y_2 , y_3 ] =
[ 1 , 0 , - i ] ; \\
q' : & (t , \lambda ) = ( 0 , \frac{1}{k'} ) &
\longleftrightarrow & [ y_1 , y_2 , y_3 ] =
[ 1 , 0 , i ] .
\end{array}
$$
The Boltzmann weights $a, b, c, d$ in Ising model can
be regarded as sections on $W_{k^\prime}$ \cite{R92}:
$$
\begin{array}{ll}
a \in \Gamma ( W_{k^\prime} , {\cal O}(q) ) , &
b \in \Gamma ( W_{k^\prime} , {\cal O}(q') ) , \\
c \in \Gamma ( W_{k^\prime} , {\cal O}(p') ) ,&
d \in \Gamma ( W_{k^\prime} , {\cal O}(p ) ) .
\end{array}
$$
Over a 4-fold cover $\widetilde{W}_{k^\prime}$ of
$W_{k^\prime}$, the above $a, b, c, d$
are linearly equivalent and satisfy the quardratic relations,
which give rise to the equations of $\widetilde{W}_{k^\prime}$ in
$\PZ^3$:
\be
\widetilde{W}_{k^\prime} : \ \left\{
\begin{array}{l}
a^2 + k' b^2 = k d^2  \\
k' a^2 + b^2 = k c^2
\end{array} \right. \ \ \ \mbox{for} \ [a,b,c,d] \in \PZ^3 ,
\ele(abcd1)
equivalently,
\be
\left\{
\begin{array}{l}
k a^2 + k' c^2 = d^2  \\
k b^2 + k' d^2 = c^2
\end{array} \right. \ \ \ \mbox{for} \ [a,b,c,d] \in \PZ^3 .
\ele(abcd2)
It is known that the variables $a, b, b, d$ have the Jacobi
elliptic function parametriaztion \cite{B} \cite{TF79}.
Here we follow the formulation in \cite{Bax} \cite{R92} by
expressing those variables via the prime form
$\vartheta_1 ( {\rm z} ) ( = \vartheta_1 ( {\rm z} , \tau ))$ of
the elliptic curve.
By formulae in \cite{R92} pp. 632- 633
\footnote{A constant was missing in the  expression of $k'$
in \cite{R92} pp. 632. The correct formula for $k'$ is as follows:
\[
k' = \frac{- e^{-\pi i ( \rho_1 + \ldots+ \rho_g )} \vartheta
[ \begin{array}{c} \bar{\delta} \\ \bar{\nu}
\end{array} ] ( \varepsilon , \tau )^N }
{i^g \ \vartheta
[ \begin{array}{c} \bar{\delta} \\ \bar{\nu}
\end{array} ] ( \varrho , \tau )^N.} .
\]
 },
together with
(\req(0THe)), $a, b, c, d$ have the following expression:
\[
a^2 : b^2 : c^2 : d^2 =   - e^{2 \pi i z}
\vartheta_1 ( {\rm z} +\frac{\tau - 1}{4})^2
: - \vartheta_1 ( {\rm z} +\frac{- \tau + 1}{4})^2
: \vartheta_1 ( {\rm z} +\frac{-\tau - 1}{4})^2
: e^{2 \pi i {\rm z}}
\vartheta_1 ( {\rm z} +\frac{\tau + 1}{4})^2
\]
with
\bea(llcl)
k' & = & - i
\frac{
\vartheta_2 ( 0 , \tau )^2 }{
\vartheta_4 ( 0 , \tau )^2 } & ( = \frac{ - e^{- \pi i \tau / 2}
\vartheta_1 ( \frac{1}{2} , \tau )^2 }{ i \
\vartheta_1 ( \frac{\tau}{2} , \tau )^2 } ) \ , \\
k & = &  \frac{
\vartheta_3 ( 0 , \tau )^2 }{
\vartheta_4 ( 0 , \tau )^2 } \ & .
\elea(ktheta)
Within constant factors, the variables $a, b, c, d$ are
proportional to the four Jacobi functions
$\vartheta_2, \vartheta_4$, $ \vartheta_3, \vartheta_1$ with
the same argument. In fact by using (\req(STTS)), one has the
following expression:
\[
a^2 : b^2 : c^2 : d^2 =    i
\vartheta_2 ( {\rm z} , \tau )^2
: \vartheta_4 ( {\rm z}  , \tau )^2
: \vartheta_3 ( {\rm z} , \tau )^2
: - i \vartheta_1 ( {\rm z} , \tau )^2  .
\]
Then the equations (\req(abcd1)) (\req(abcd2)) are equivalent
to the 2nd-4th and 3rd-1st relations in (\req(relth)) for Jacobi elliptic
functions.
By Sect. 3 of \cite{R92}, the variables $\lambda , t$ are
related to $a, b, c, d$ by
\[
\lambda = \frac{d^2}{c^2} \ , \ \ \ t = \frac{ab}{cd} \ ,
\]
hence we obtain theta function representations of $\lambda$
and $ t$:
\bea(lll)
\lambda = & \frac{ e^{2 \pi i {\rm z}}
\vartheta_1 ( {\rm z} + \frac{\tau + 1}{4} , \tau )^2 }
{ \vartheta_1 ( {\rm z} - \frac{\tau + 1}{4} , \tau )^2 } &
= - i \frac{
\vartheta_1 ( {\rm z}  , \tau )^2 }
{ \vartheta_3 ( {\rm z}  , \tau )^2 } \ , \\ [2mm]
t = &
\frac{
\vartheta_1 ( {\rm z} + \frac{\tau - 1}{4} , \tau )
\vartheta_1 ( {\rm z} + \frac{- \tau + 1}{4} , \tau ) }
{ \vartheta_1 ( {\rm z} - \frac{\tau + 1}{4} , \tau )
\vartheta_1 ( {\rm z} + \frac{\tau + 1}{4} , \tau )
} & = - i
\frac{
\vartheta_2 ( {\rm z}  , \tau )
\vartheta_4 ( {\rm z}  , \tau ) }
{ \vartheta_3 ( {\rm z}  , \tau )
\vartheta_1 ( {\rm z} , \tau ) } \ .
\elea(tlambda)

Note that the Picard-Fuchs equation for the Ising-family
(\req(Ising)) is equivalent to that of $X_9$-family. In fact,
the equation is derived
by Dwork-Griffith-Katz reduction method
from residuum expression of the period:
\[
\hat{\omega}( s_0, s_1, s_2) = \int_{\gamma} \int_{\Gamma_i}
\frac{y_1 dy_2 \wedge dy_3 - y_2 dy_1 \wedge dy_3 +
\frac{1}{2} y_3 dy_1 \wedge dy_2}
{s_1 y_1^4 + s_2y_2^4 + y_3^2 - s_0 y_1^2 y_2^2} \ ,
\]
where $\gamma$ is a small circle in $\PZ^2_{(1,1,2)}$ normal to
the elliptic curve ,
$\Gamma_i $ are 1-circles on the curve. The above integral
is also expressed by:
\[
\hat{\omega}( 1, 1, \epsilon ) =
\frac{-1}{2} \int_{\Gamma_i}
\frac{dt}{w} \ , \ \ \
\epsilon = 2 \frac{1+k'^2}{1-k'^2}
\]
where $(t, w)$ are the coordinates of $W_{k'}$ in (\req(eqnw)).
It is known that
$\hat{\omega}( s_0, s_1, s_2)$ satisfies the folllowing
equations:
\be
\left\{ \begin{array}{ll}
( s_0 \frac{ \partial }{ \partial s_0 } +
s_1 \frac{ \partial }{ \partial s_1 } +
s_2 \frac{ \partial }{ \partial s_2 } + \frac{1}{2} ) \hat{\omega}
& = 0 \ ,   \\
( s_1 \frac{ \partial }{ \partial s_1 } -
s_2 \frac{ \partial }{ \partial s_2 } ) \hat{\omega}
& = 0 \ , \\
( \frac{ \partial }{ \partial s_1 } \frac{ \partial }
{ \partial s_2 } -
 \frac{ \partial^2 }{ \partial s_0^2 }  ) \hat{\omega}
& = 0 \ .
\end{array}
\right.
\ele(PFe)
By the ansatz
\[
\hat{\omega}( s_0, s_1, s_2) =
\frac{1}{\sqrt{s_0}} \omega ( \zeta ) \ , \ \ \
\zeta : = \frac{s_1 s_2}{s_0^2} = \frac{1}{\epsilon^2}\ ,
\]
the equation (\req(PFe)) is brought into the form
\be
[ 4 ( 4 \zeta - 1 ) ( \zeta \frac{ \partial  }{ \partial
\zeta } )^2 +
16 \zeta ^2 \frac{ \partial }{ \partial \zeta } + 3
\zeta ] \omega ( \zeta )  = 0 \ ,
\ele(eqzeta)
which has 3 regular singular points at
\[
\zeta = 0 , \ \infty , \ \frac{1}{4} \ .
\]
By the change of coordinates,
\[
z = \frac{- \zeta}{16} + \frac{1}{64} =
\frac{- 1}{16 \epsilon^2 } + \frac{1}{64} \ ,
\]
the equation (\req(eqzeta)) becomes
\[
[ ( z \frac{ \partial }{ \partial z } )^2 - 4
(4 z \frac{ \partial }{ \partial z } + 3 )
(4 z \frac{ \partial }{ \partial z } + 1 ) ] \omega ( z ) = 0 ,
\]
which is the differential opereator in Table (I) for $X_9$.

\section{Jacobi Elliptic Function Parametrization of
$X_9$-family }
In this section we investigate the $X_9$-family :
$$
\begin{array}{lll}
( X_9 ) \ \ \ \ \  & \ X_s :  f_s ( x ) =
x_1^4 + x_2^4 +x_3^2 - s x_1 x_2 x_3 = 0 , &
[x_1, x_2, x_3 ] \in {\PZ}^2_{(1,1,2)} .
\end{array}
$$
The curve $X_s$ degenerates at
\[
s = \infty  \ , \ \ 2 \sqrt{2} \omega \ \ ( \ \omega^4 = 1 ) \ ,
\]
and for $s = 2 \sqrt{2} \omega$,  it becomes the
union of two rational curves:
\[
x_1^4 + x_2^4 +x_3^2 - 2 \sqrt{2} \omega x_1 x_2 x_3  =
( x_3 - \sqrt{2} \omega x_1 x_2 )^2 +
( x_1^2 - \omega^2 x_2^2 )^2 = 0 \ .
\]
The linear transformation group
preserving a gereric polynomal $f_s ( x )$ is generated
by $C, R, \Sigma$ which are defined by
\bea(ll)
C ( x_1, x_2 , x_3 ) =
( i x_1 , i x_2 , - x_3 ) \ , \ &
R ( x_1, x_2 , x_3 ) = ( - x_1 , x_2 , -x_3 ) \ , \\
\Sigma ( x_1, x_2 , x_3 ) = ( x_2 , x_1 , x_3 ) \ \ . &
\elea(CRI)
The subgroup generated by $C$ and $ R$,
\[
< C , R > = \{ \mbox{dia.} ( \alpha_1, \alpha_2, \alpha_3 ) \
| \ \alpha_1^4 =
\alpha_2^4 = \alpha_3^2 =
\alpha_1 \alpha_2 \alpha_3 = 1 \} \ \simeq \
\ZZ_4 \oplus \ZZ_2 ,
\]
acts on ${\PZ}^2_{(1,1,2)}$ as an order 2 group induced by $R$.
Denote $r_s , \widetilde{\sigma}_s$ the restriction of $R$
and $\Sigma$ on $X_s$,
\[
r_s : X_s \longrightarrow X_s \ , \ \ \
\widetilde{\sigma}_s : X_s \longrightarrow X_s \ .
\]
Then $r_s$ is an order 2 translation and $\widetilde{\sigma}_s$ is an
involution
of $X_s$ with
\[
r_s \cdot \widetilde{\sigma}_s =
\widetilde{\sigma}_s \cdot r_s \ .
\]
The zeros of $x_i$'s in $X_s$ are given by
\[
 \mbox{zero} ( x_1 ) = \{  [ 0 , 1 , \pm 1 ] \} \ ; \ \
 \mbox{zero} ( x_2 ) = \{  [ 1 , 0 , \pm 1 ] \} \ ; \ \
 \mbox{zero} ( x_3 ) =
\{  [ 1 , \eta , 0 ] \ , \eta^4 = -1 \} ,
\]
each of which is stable under the action of $r_s$.
Via the the projection
\[
p : {\PZ}_{(1,1,2)}^2  \longrightarrow
{\PZ}_{(1,1,2)}^2 /<R> \ ,
\]
the coordinates $x_i$'s of ${\PZ}_{(1,1,2)}^2$ give rise to
sections $\xi_i$'s on ${\PZ}_{(1,1,2)}^2/<R> $ with
$p^*( \xi_i )= x_i$.
We have a family of elliptic curves in
${\PZ}_{(1,1,2)}^2/<R>$ :
$$
\begin{array}{lll}
 \Xi_s :&
\xi_1^4 + \xi_2^4 + \xi_3^2 - s \xi_1 \xi_2 \xi_3 = 0 , &
[\xi_1, \xi_2, \xi_3 ] \in {\PZ}^2_{(1,1,2)}/<R> .
\end{array}
$$
The restriction of $p$ defines a 2-fold cover of elliptic curves:
\[
p_s : \ X_s \longrightarrow \Xi_s : = X_s/<r_s> \ .
\]
Let $\sigma_s$ be the involution of $\Xi_s$ induced by
$\widetilde{\sigma}_s$,
$$
\begin{array}{clc}
X_s & \stackrel{\widetilde{\sigma}_s}{\longrightarrow}& X_s
\ \ \\
\downarrow & & \downarrow \ \ \\
\Xi_s & \stackrel{\sigma_s}{\longrightarrow}& \Xi_s \ ,
\end{array}
$$
and denote the zeros of $\xi_i$'s in $\Xi_s$ by
\[
 \mbox{zero} ( \xi_1 ) = \{  p \} \ ; \ \
 \mbox{zero} ( \xi_2 ) = \{  p^\prime \} \ ; \ \
 \mbox{zero} ( \xi_3 ) =
\{  p_3 , p_4 \} .
\]
We have
\[
2 p  \sim 2 p^\prime \ , \
p + p^\prime \sim p_3 + p_4 \ \ ;  \ \ \
\mbox{and} \ \ \ \ \xi_1^2 , \xi_2^2 \in
\Gamma (\Xi_s , {\cal O} ( 2 p ) ) \ , \ \ \
\xi_1\xi_2  , \ \xi_3
\in \Gamma ( \Xi_s , {\cal O} ( p_3 + p_4 ) ) \ .
\]
The elements $p, p', p_3, p_4$ determine the equation of
$\Xi_s$ by the following
lemma. However as we shall see it later on, much efforts are required
in order to obtain some theta function forms of $\xi_i$'s.
 \par \vspace{0.2in} \noindent
{ \bf Lemma 5.} Let $\EZ$ be an elliptic curve with two elements
$p , p'$ in it fixed by an involution $\theta$.
Let $m$ be the order 2 translation of $\EZ$ with
$m ( p ) = p' $. Then

(i) There exist sections
\[
f_1 \in \Gamma ( \EZ , {\cal O}(p) ) \ , \ \
f_2 \in \Gamma ( \EZ , {\cal O}(p') )
\]
such that the following diagram commutes:
$$
\begin{array}{ccccc}
\EZ & \simeq & \EZ & \stackrel{m}{\longrightarrow} & \EZ \\
\downarrow & & \downarrow \Psi & & \downarrow \Psi \\
\EZ / <\theta> & \simeq & \PZ^1 &\stackrel{m_0}{\longrightarrow}
& \PZ^1 \ ,
\end{array}
$$
where $\Psi ( x ) = [ f^2_1 ( x ) , f^2_2 ( x ) ]$ and
$m_0 ( [\xi, \eta ] ) = [\eta, \xi ] $.

(ii) Let $p_3$ be an element of $\EZ$ with $
\Psi ( p_3 ) = [ 1 , i ]$ , and $
p_4 := m ( \theta ( p_3 ) ) $.
Then $p_3 \neq p_4 $ ,
\[
 4 p \sim 4 p' \sim 2 p_3+ 2 p_4 \sim
p + p' + p_3 + p_4 ,
\]
and for some $ s \in \CZ$ and $
f_3  \in \Gamma ( \EZ , {\cal O}(p_3 + p_4) ) $ with
$\mbox{zero} ( f_3 ) = p_3 + p_4$, the following relation
holds :
\[
f_1^4  + f_2^4 + f^2_3 = s f_1 f_2 f_3 \ \in
\Gamma ( \EZ , {\cal O}(4 p ) ) \ .
\]
\par \vspace{0.2in} \noindent
{\it Proof.}
By identifying $\EZ/ <\theta>$ with $\PZ^1$,
we may assume the map
\[
\Psi : \EZ \longrightarrow \PZ^1 = \EZ / <\theta>
\]
sends $p , p'$ to $[0, 1] , [1, 0]$ respectively.
Since $m$ commutes with $\theta$, there is an order 2 automorphism
$m_0$ of $\PZ^1$ which interchanges the elements $ [0, 1]$ and
$ [1, 0]$,
hence for some coordinate system of $\PZ^1$,
$m_0$ is described by
\[
m_0 ( [\xi, \eta ] ) = [\eta, \xi ] .
\]
Then the sections $f_1 , f_2$ in (i) are easily obtained .
For $x, y \in \EZ$, ons has
\[
y = m ( x ) \Longleftrightarrow x + p' \sim p + y \ .
\]
By the definition of $p_3 $ and $ p_4$, we have the linearly
equivalent relations:
\[
2 p + p' \ \sim \
p_3 + \theta ( p_3 ) + p' \ \sim \ p + p_3 + p_4 \ ,
\]
hence
\[
p + p' \sim  p_3 + p_4\ .
\]
Then the equivalent relations in (ii)
follow immediately. By $\Psi (  \theta ( p_3 ) ) =
\Psi (  p_3 ) $, we
have
\[
\Psi ( p_4 ) = \Psi ( m ( \theta ( p_3 ) ) ) =
m ( \Psi (  \theta ( p_3 ) ) ) =
m ( [ 1 ,  i ] )
 = [ 1 , - i ] \ .
\]
Therefore $ p_3 \neq p_4$, and
\[
f_1^4 ( p_j ) + f_2^4 ( p_j ) = 0 \ \ \mbox{for} \ j = 3 , 4 .
\]
Let $f_3$ be a section in $\Gamma ( \EZ , {\cal O}(p_3 + p_4) ) $
with
\[
\mbox{zero} ( f_3 ) = p_3 + p_4 \ , \ \ \ \
f_2^4 ( p ) + f_3^2 ( p ) = 0 \ .
\]
Both sections $f_1^4 + f_2^4 + f_3^2$ and
$f_1f_2f_3$ in $\Gamma ( \EZ , {\cal O}(4p) )$ vanish at
$p_3 , p_4$ and $p$, hence they are proportional by a
non-zero constant. Therefore we obtain (ii).
$\Box$ \par \vspace{0.2in} \noindent

Consider the birational map
\[
\phi : {\PZ}_{(1,1,2)}^2 \longrightarrow \PZ^1 \times \PZ^1 \ ,
\ \ [x_1, x_2, x_3] \mapsto ([ x_1 , x_2 ] , [x_3 , x_1x_2] )
\ ,
\]
whose fundamental locus
consists of 3 elements, defined by two of coordinates
$x_i$'s being zero. Outside the fundamental locus, $\phi$  defines
an isomorphism. The morphism $R$ of ${\PZ}_{(1,1,2)}^2$ in
(\req(CRI)) induces
the morphism $\widetilde{R} $,
\[
\widetilde{R} : \PZ^1 \times \PZ^1
\longrightarrow \PZ^1 \times \PZ^1 \ , \ \
( [y_1 , y_2], [y_3 , y_4] ) \mapsto
( [y_1 , -y_2], [y_3 , y_4] ) \ ,
\]
with the commutative diagram:
$$
\begin{array}{clc}
{\PZ}_{(1,1,2)}^2 & \stackrel{\phi}{\longrightarrow}
& \PZ^1 \times \PZ^1 \ \ \\
R \downarrow \ \ \ \ \ \ \ & & \downarrow \widetilde{R}  \\
{\PZ}_{(1,1,2)}^2 & \stackrel{\phi}{\longrightarrow}
& \PZ^1 \times \PZ^1 \  .
\end{array}
$$
Hence $\phi$ induces a birational morphism between
 ${\PZ}_{(1,1,2)}^2 / <R> $ and $\PZ^1 \times \PZ^1$:
\[
 {\PZ}_{(1,1,2)}^2/ <R> \ \ \longrightarrow \
\PZ^1 \times \PZ^1 \ ,
\ \ [\xi_1, \xi_2, \xi_3] \mapsto ([ \xi^2_1 , \xi^2_2 ] ,
[\xi_3 , \xi_1\xi_3] ) \ ,
\]
under which $\Xi_s$ is embedded into
$\PZ^1 \times \PZ^1$. The coordinates of $\Xi_s$ in
$\PZ^1 \times \PZ^1$ are now given by
\bea(lll)
\Psi : \Xi_s \longrightarrow \PZ^1  \ , &
x \mapsto [\xi_1^2 ( x ) , \xi_2^2 ( x ) ] \ ,
& \lambda ( x ) : =  \frac{\xi_1^2}{\xi_2^2} ( x ) \ ; \\ [2mm]
\Pi^\prime : \Xi_s \longrightarrow \PZ^1  \ , &
x \mapsto [\xi_3 ( x ) , \xi_1 \xi_2 ( x )  ] \ ,
& u ( x ) : =  \frac{\xi_3}{\xi_1 \xi_2} ( x ) \ ,
\elea(PsiWs)
which defines a plane curve birational to $\Xi_s$:
\[
u^2 - s u = - ( \lambda + \frac{1}{\lambda} ) \ , \ \ \
( u , \lambda ) \in \CZ^2 \ .
\]
By the relation
\[
\Pi^\prime \sigma_s = \Pi^\prime
\]
and the commutative diagram:
$$
\begin{array}{ccc}
\Xi_s & \stackrel{\sigma_s}{\longrightarrow} & \Xi_s  \\
\Psi \downarrow \ \ \ \ &  & \Psi \downarrow \ \ \ \ \\
\PZ^1 & \stackrel{\lambda \mapsto
\lambda^{-1}}{\longrightarrow} & \ \PZ^1 \ ,
\end{array}
$$
the morphism $\Pi^\prime$ is equivalent to the projection,
\[
\Xi_s \longrightarrow \Xi_s/<\sigma_s> \ ,
\]
and for some involution $\theta_s$ of $\Xi_s$, the following
relstions hold:
\[
\Psi \theta_s = \Psi \ , \ \
 \theta_s \sigma_s = \sigma_s \theta_s \ .
\]
It is easy to see that the branched data of $\Psi$ and
$\Pi^\prime$ are given by
$$
\begin{array}{ll}
\mbox{Branched \ points \ of\ } \Psi :&
( u , \lambda ) = ( \frac{s}{2} , k' ) , \
(\frac{s}{2} , \frac{1}{k^\prime} ) , \
( \infty , 0) , \  ( \infty , \infty ) \ ;   \\
\mbox{Branched \ ponits \ of \ } \Pi^\prime :&
( u , \lambda ) = ( \frac{ s \pm \sqrt{s^2 - 8} }{2}, 1 ) \ ,
\ ( \frac{ s \pm \sqrt{s^2 + 8}}{2} , -1 ) \ ,
\end{array}
$$
where $k'$ is defined by
\[
k^\prime + \frac{1}{k^\prime} = \frac{s^2}{4} \ .
\]
Change the variable of $\PZ^1$ from $u$ to $t$ via
\be
t = \frac{2}{\sqrt[4]{s^4 - 64 }} ( u - \frac{s}{2} ) \
\ ( \ = \sqrt{ \frac{k'}{k^2}} [ \
u - (k^\prime + \frac{1}{k^\prime} )^{1/2} ] ) \ ,
\ele(turel)
and define the morphism
\be
\Pi : \Xi_s \longrightarrow \PZ^1 \ , \ \
\Pi ( x ) : = t ( \Pi^{\prime} ( x ) ) \ .
\ele(Pit)
The branched locus $\Pi$ becomes
\[
t = \pm \sqrt{ \frac{1-k'}{1+k'}} \ , \ \
\pm \sqrt{ \frac{1+ k'}{1-k'}} \ .
\]
With $\Psi$ in (\req(PsiWs)) and $\Pi$ in (\req(Pit)), one may
identify $\Xi_s$ with the curve $W_{k^\prime}$ (\req(N=2)),
which is the same as $Y_{\epsilon}$ (\req(Ising)).
By (\req(PsiWs)) (\req(turel)) and (\req(twyi)), the following
elliptic curves are isomorphic:
\bea(ccccc)
\Xi_s & \simeq & W_{k^\prime} & \simeq & Y_{\epsilon}  \\
\ [ \xi_1, \xi_2, \xi_3 ] & \longleftrightarrow & ( t , \lambda )
& \longleftrightarrow & [y_1, y_2, y_3] \\
\elea(XiWY)
with the correspondences:
$$
\begin{array}{cll}
( t , \lambda  ) & = &
(  \frac{2\xi_3 - s \xi_1 \xi_2}
{\sqrt[4]{s^4 - 64 } \xi_1 \xi_2} ,
\ \frac{\xi_1^2}{\xi_2^2} ) \\ [2mm]
\ [ y_1 , y_2 , y_3 ] & = & [ \frac{\sqrt{s^4 - 64 }}{2}
\xi_1 \xi_2 , \ \xi_3 - \frac{s}{2}\xi_1 \xi_2 , \
i ( \xi_1^4 - \xi_2^4 ) ] \ ,
\end{array}
$$
where parameters $s, k^\prime$ and $\epsilon$ are related by
\[
 \frac{s^2}{4} = k^\prime + \frac{1}{k^\prime} \ , \ \
\epsilon = 2\frac{1+ k'^2}{1-k'^2} =
\frac{2s^2}{\sqrt[4]{s^4 - 64 }}\ .
\]
Note that
for the $X_9$-family $\{ X_s \}_s$ and the Ising family $\{
W_{k^\prime} \}_{k^\prime}$ (\req(Ising)), one has
$$
\begin{array}{cll}
X_s \simeq X_{s_1} & \Longleftrightarrow & s_1 = \omega s \
\ \ \ \ ( \ \omega^4 = 1 \ ) \ ; \\
W_{k^\prime}  \simeq W_{k_1^\prime}  &
\Longleftrightarrow & k_1^\prime = \pm k^\prime ,
\pm k^{\prime -1} \ .
\end{array}
$$
Hence the variable $z$,
\be
z : = s^{-4} = \frac{-1}{16 \epsilon^2} + \frac{1}{64} =
\frac{ k'^2 }{ 16 ( 1 + k'^2 )^2 } \ ,
\ele(zk)
is considered as the moduli parameter of the isomorphic classes
of elliptic curves
either in $X_9$-family, or in Ising family. According to the
discussion in Sect. 5, one may identify $W_{k^\prime}$ with
$\EZ_{\tau, 1}$ where $k^\prime , \tau$ satisfy the relation
(\req(ktheta)). Using (\req(0THe)), we have
\be
z ( \tau )  =
\frac{ - \vartheta_2 (0 , \tau )^4 \vartheta_4 (0 , \tau )^4}
{16 ( \vartheta_3 (0 , \tau )^8 -
4 \vartheta_2 (0 , \tau )^4 \vartheta_4 (0 , \tau )^4 ) } ,
\ele(zX9)
hence
\[
s ( \tau )  =  2 e^{\frac{3 \pi i}{4}} \frac{\sqrt
{\vartheta_2 (0 , \tau )^4 - \vartheta_4 (0 , \tau )^4} }
{\vartheta_2 (0 , \tau )\vartheta_4 (0 , \tau )} \ , \ \
\ \ \ \
\frac{\sqrt[4]{s(\tau)^4 - 64 }}{2}  =  e^{\frac{ \pi i}{4}}
\frac{\vartheta_3 (0 , \tau )^2}
{\vartheta_2 (0 , \tau )\vartheta_4 (0 , \tau )} \ .
\]
{ \bf Theorem 4. } With $x_i ,
\xi_i \ ( 1 \leq i \leq 3)$ the coordinates of
elliptic curves $X_s , \Xi_s$ in the $X_9$-family, $p_s$ the morphism between
them, and $ r_s , \widetilde{\sigma}_s  $ the automorphisms of $X_s$
as before.
For $\tau \in \HZ$, define $s ( \tau )$ by
\[
s ( \tau )  =  2 e^{\frac{3 \pi i}{4}} \frac{\sqrt
{\vartheta_2 (0 , \tau )^4 - \vartheta_4 (0 , \tau )^4} }
{\vartheta_2 (0 , \tau )\vartheta_4 (0 , \tau )} .
\]
Then the above data for $X_s , \Xi_s$ have the
following realization in complex tori:
$$
\begin{array}{lll}
X_s = \EZ_{ \tau + 1 , 2}, & \Xi_s = \EZ_{ \tau , 1} , &
 p_s : X_s \longrightarrow \Xi_s \ , \ [{\rm z}] \mapsto
[{\rm z}] \ ;
\\
r_s : \EZ_{ \tau +1 , 2} \longrightarrow \EZ_{ \tau +1 , 2} ,&
 [{\rm z}] \mapsto [{\rm z} + 1]  \ ; \\
\widetilde{\sigma}_s : \EZ_{ \tau +1 , 2} \longrightarrow
\EZ_{ \tau +1 , 2} ,&
[{\rm z}] \mapsto [-{\rm z} + \frac{\tau+1}{2}] ,
\end{array}
$$
with $\xi_i$'s given by
$$
\begin{array}{lll}
\xi_1  & = &  \vartheta_1 ( {\rm z} , \tau ), \\
\xi_2 &  = & e^{\frac{\pi i}{4}}
\vartheta_3 (  {\rm z} , \tau ) , \\
 \xi_3 & =  &\frac{\vartheta_3 (0 , \tau )^2}
{\vartheta_2 (0 , \tau )\vartheta_4 (0 , \tau )}
\vartheta_2 ( {\rm z} , \tau ) \vartheta_4 ( {\rm z} , \tau )
- \frac{\sqrt
{\vartheta_2 (0 , \tau )^4 - \vartheta_4 (0 , \tau )^4}}
{\vartheta_2 (0 , \tau )\vartheta_4 (0 , \tau )}
\vartheta_1 ( {\rm z} , \tau ) \vartheta_3 ( {\rm z} , \tau )
\ ,
\end{array}
$$
and through the Heisenberg group action on entire functions,
one has the projective representation of $< r_s ,
\widetilde{\sigma}_s > $ on
$x_i$'s :
$$
\begin{array}{llll}
r_s :& (x_1, x_2 , x_3 ) & \mapsto & ( - x_1 , x_2 , x_3 ) , \\
\widetilde{\sigma}_s  :& (x_1, x_2 , x_3 ) & \mapsto &
( - e^{\pi i/4} x_2 , - e^{\pi i/4} x_1 , e^{\pi i/2}x_3 ) .
\end{array}
$$
\par \vspace{0.2in} \noindent
{\it Proof.} According to the discussion we have before, with
$s = s ( \tau)$ one has the identification:
\[
\Xi_s = \EZ_{ \tau , 1} \ ,
\]
and the rational functions $\Psi, \Pi$ in (\req(PsiWs))
(\req(Pit) ) equal to
$$
\begin{array}{lll}
\Psi :& \EZ_{ \tau , 1} \longrightarrow \PZ^1 =
\EZ_{ \tau , 1}/<\theta> \ , & \mbox{where} \ \ \
\theta ( [ {\rm z} ] ) = [ -{\rm z}] \ ; \\
\Pi : & \EZ_{ \tau , 1} \longrightarrow \PZ^1 =
\EZ_{ \tau , 1}/<\sigma> \ , &\mbox{where} \ \ \
\sigma ( [ {\rm z} ] ) = [ -{\rm z} + \frac{\tau+1}{2}] \ .
\end{array}
$$
Note that the above $\sigma$ is identified with the
automorphism $\sigma_s$ of $\Xi_s$, which can be lifted to the
automorphism $\widetilde{\sigma}_s$ on $X_s$.
Write $X_s = \CZ / L$ for some index 2
sublattice $L$ of $\ZZ \tau + \ZZ$. The morphism $p_s$ is
given by the natural projection. On the univeral covering
space $\CZ$ of $X_s$, the affine map
\[
{\rm z} \mapsto - {\rm z} + \frac{\tau+1}{2} \ , \ \
{\rm z} \in \CZ \ ,
\]
induces the order 2 automorphism $\widetilde{\sigma}_s$ on
$X_s$, hence the element $\tau + 1 $ is in the lattice $L$. This implies
$X_s = \EZ_{ \tau + 1 , 2}$ with $r_s , \widetilde{\sigma}_s $
 described in the theorem. The Jacobi elliptic function
parametrization of $\xi_i$'s now follows from (\req(tlambda))
(\req(XiWY)), and the action of $r_s ,
\widetilde{\sigma}_s $ on $x_i$'s is obtained by formulae in
Sect. 2.
$\Box$ \par \vspace{0.2in} \noindent
Now the formula
for $z \ ( = s^{-4}) $ with $s = s ( \tau)$ can be derived from
 the above theorem as follows:
\par \vspace{0.2in} \noindent
{\bf Theorem 5 .} The function $z ( {\bf q} )$ for
$X_9$-family  of Table (I) is given by
$$
\begin{array}{lll}
z ( {\bf q } ) & =
\frac{ - \vartheta_2 (0 , 2{\bf t}-1 )^4
\vartheta_4 (0 , 2{\bf t}-1 )^4 }
{ 16 \vartheta_3 (0 , 2{\bf t}-1 )^8 -
64 \vartheta_2 (0 , 2{\bf t}-1 )^4
\vartheta_4 (0 , 2{\bf t}-1 )^4  } & \\ [2mm]
 & = \frac{ {\bf q} \prod_{n=1}^{\infty} ( 1 + {\bf q}^n )^8}
{\prod_{n=1}^{\infty} ( 1 - {\bf q}^{2n-1} )^{16} + 64
{\bf q} \prod_{n=1}^{\infty} ( 1 + {\bf q}^n )^8 } \ ,& \
{\bf q} = e^{2 \pi i {\bf t}} \ .
\end{array}
$$
As a consequence, $z ({\bf q})$ has an integral
${\bf q}$-expansion with
\[
\lim_{{\bf q} \rightarrow 0 }
\frac{z ({\bf q} )}{ {\bf q}} = 1 \ .
\]
\par \vspace{0.2in} \noindent
{\it Proof. }
For $s = s ( \tau )$ in Theorem 4,
we have
\[
X_s = \EZ_{ \tau + 1 , 2} \simeq \EZ_{ {\bf t} , 1} \ , \
\  {\bf t} = \frac{\tau+ 1}{2} \ .
\]
With the theta-constant expression (\req(ktheta)) for
$k^{\prime}$, the relation (\req(zk)) gives rise the expression
of $z ( {\bf q})$. The same argument as in Theorem 3 shows that
${\bf t}$ is the variable described in Sect. 3. Therefore we have
completed the proof of this theorem.
$\Box$ \par \vspace{0.2in} \noindent

\section{Discussion}
In this paper, we have focused on the mathematical structure
of  "counting" functions $z ({ \bf q})$ , and have developed
an analysis of constrained
Table (I) by means of elliptic theta function representations.
Now we are going to discuss another aspect, which is
possibly of certain physical relevance.
Here the explictly work out example of $X_9$-family has illustrated
its close connection with Ising model in statistical
mechanics, where we employ the Jacobi elliptic function
parametrization of Ising model to the investigation of
$X_9$-potential. The relation (\req(zk)) states the parameter
$z$ in $X_9$-family corresponds to the
temperature-like parameter $k^\prime$ of Ising model.
One interesting point for the derivation of the function
$z ({ \bf q})$ is that on one hand it is related to
Picard-Fuschs equation of the elliptic family, while on the other
side with
the parametrization for
Boltzmann weights of the statistical model,
the same result is correctly reproduced. In this setting, the
mirror symmetry of $X_9$-family is connected to the $\ZZ_2$-cover of
elliptic curves in Ising model, which often appears in the
theory.
Due to the relative simplicity of the models, the quantities
involved in our mathematical work usually have
interpretations of physical or geometrical meaning, which
allows one to compare of their essential structures
in an explicit way:
$$
\begin{array}{clc}
\underline{\rm N=2 \ SUSY \ LG \ theory} &  &
\underline{\rm 2-dim. \ exactly \ solvable \  model } \\ [4mm]
{\rm LG \ fields} & \longleftrightarrow &
{\rm Boltzmann \ weights } \\
{\rm LG \ superpotential} & \longleftrightarrow &
{\rm Yang-Baxter \ equation } \\
{\rm moduli \ parameter} & \longleftrightarrow &
{\rm temperature-like \ parameter } \\
{\rm maximally \ unipotent \ area} & \longleftrightarrow &
{\rm low \ temperature \ region} \\
\end{array}
$$
Though we do not know now whether other models are related in
a similar manner, it should be interesting to note that in
the work carried out in this article,
the mathematical structures of corresponding concepts
do share a common feature. The theta function
parametrization of Ising model we used here has a
direct generalization to chiral Potts $N$-state models.
The naive quantitative indications presented by two physical
theories, $X_9$ and Ising models, encourage us to seek a
possible link between Calabi-Yau manifolds and chiral Potts
models.
$$
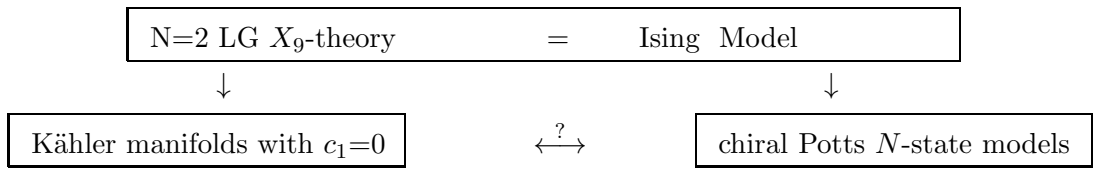
\put(160, 20){\line(0, -1){20}}
\put(-155, 20){\line(1, 0){315}}
\put(-155, 0){\line(1, 0){315}}
\put(-155, 20){\line(0, -1){20}}
\put(-150, 5){ \shortstack{ N=2 LG $X_9$-theory}}
\put(35, 5){ \shortstack{ Ising \ Model }}
\put(0, 5){ \shortstack{ = }}
\put(-200, -20){\line(0, -1){20}}
\put(-200, -20){\line(1, 0){150}}
\put(-200, -40){\line(1, 0){150}}
\put(-50, -20){\line(0, -1){20}}
\put(-195, -35){ \shortstack{ K\"{a}hler manifolds with $c_1$=0 }}
\put(-125, -12){ \shortstack{ $\downarrow$}}
\put(60, -20){\line(0, -1){20}}
\put(60, -20){\line(1, 0){150}}
\put(60, -40){\line(1, 0){150}}
\put(210, -20){\line(0, -1){20}}
\put(65, -35){ \shortstack{ chiral Potts $N$-state models}}
\put(105, -12){ \shortstack{ $\downarrow$ }}
\put(-5, -35){ $\stackrel{?}{ \longleftrightarrow}$ }
$$
The connection proposed in above diagram is vague. Neverless some
of symmetries presented in the study of Calabi-Yau spaces resemble those in
chiral Potts $N$-state models. So, we hope some appropriate geometric
picture does exist. How to detect this novel phenomena should be
of merit for further investigation.
\par \vspace{0.4in} \noindent
{\bf Acknowledgements} \par \noindent
The author would like to thank A. Klemm, B. H. Lian,
and S. T. Yau for discussions on mirror symmetry, and
also to S. S. Lin for discussions
on 2-dimensional solvable statistical models.

\end{document}